\documentclass[pdflatex,sn-mathphys-num]{sn-jnl}% Math and Physical Sciences Numbered Reference Style
\usepackage{graphicx}%
\usepackage{multirow}%
\usepackage{amsmath,amssymb,amsfonts}%
\usepackage{amsthm}%
\usepackage{mathrsfs}%
\usepackage[title]{appendix}%
\usepackage{xcolor}%
\usepackage{textcomp}%
\usepackage{manyfoot}%
\usepackage{booktabs}%
\usepackage{algorithm}%
\usepackage{algorithmicx}%
\usepackage{algpseudocode}%
\usepackage{listings}%
\usepackage{amsmath,amssymb,amsfonts}%
\numberwithin{equation}{section}
\def\nn{\nonumber}
\def\rnd{\partial}
\def\x{\mathrm{x}}

\begin{document}

\title[Article Title]{Induced energy-momentum tensor of the scalar field in 3D de~Sitter~QED}

\author*[1]{\fnm{Manizheh } \sur{Botshekananfard}}\email{manizheh.botshekananfard@bogazici.edu.tr}

\author[2]{\fnm{Takahiro} \sur{Hayashinaka}}

\affil*[1]{\orgdiv{Department of Physics},
\orgname{Bo\u{g}azi\c{c}i University},
\orgaddress{\street{Bebek},
\city{\.{I}stanbul},
\postcode{34342},
\country{Turkey}}}

\affil[2]{\orgdiv{Former student of Research Center for the Early Universe (RESCEU)},
\orgname{Graduate School of
Science, The University of Tokyo},
\orgaddress{\street{7-3-1 Hongo, Bunkyo-ku},
\city{Tokyo},
\postcode{113-0033},
\country{Japan}}}

\abstract{In this work, we derive the renormalized expectation value of the energy--momentum tensor of a quantized charged scalar field in three-dimensional de Sitter spacetime $\mathrm{dS}_{3}$ in the presence of a uniform electric field. Using the adiabatic regularization method, ultraviolet divergences are systematically removed, yielding finite expressions for all components of the induced tensor.
We analyze the behavior of the renormalized energy--momentum tensor in both the strong-field and infrared regimes. In the strong-field limit, the induced energy density exhibits a quadratic dependence on the electric field strength; however, this leading contribution is dominated by vacuum polarization effects rather than directly by the Bogoliubov particle-production component. In the infrared regime, the tensor shows a pronounced inverse-mass dependence, indicating strong infrared sensitivity characteristic of light scalar fields in de Sitter spacetime.
In the conformally coupled massless limit, the renormalized trace vanishes, as expected from the absence of a genuine Weyl anomaly in odd-dimensional spacetimes. These results provide a precise characterization of vacuum polarization and infrared effects in three-dimensional de Sitter scalar QED.}


\maketitle

\section{Introduction}
\subsection{Motivation and scope}
Quantum field theory in curved spacetime provides a semiclassical framework in which particle creation and vacuum polarization arise due to curvature and external fields.
The renormalized expectation value of the energy--momentum tensor 
$\langle T_{\mu\nu} \rangle_{\mathrm{ren}}$ is a central observable for characterizing quantum effects of matter fields in curved spacetime, as it encodes both vacuum polarization and state-dependent particle-production contributions and determines the backreaction of quantum fields on the background geometry within semiclassical gravity 
\cite{BirrellDavies1982,ParkerToms2009,Wald1994,PhysRevLett.21.562,PhysRev.183.1057,PhysRevD.3.346}.

In curved backgrounds, the Schwinger effect is modified by spacetime expansion, leading to a nontrivial interplay between electromagnetic and gravitational particle production.
In this work, we study a quantized charged scalar field in 
three-dimensional de Sitter spacetime ($\mathrm{dS}_3$) in the presence of 
a uniform electric field. This setup combines gravitational particle creation 
\cite{PhysRevLett.21.562,PhysRev.183.1057,PhysRevD.3.346} 
with the Schwinger effect 
\cite{PhysRev.82.664,gelis2016schwinger,PhysRevD.49.6343}, 
providing a minimal framework in which curvature and gauge-field effects are simultaneously realized.

The three-dimensional de Sitter geometry is particularly well suited for this problem. Owing to its reduced dimensionality and the absence of local gravitational degrees of freedom \cite{deser1984three,witten19882+}, 
$\mathrm{dS}_3$ allows for analytically controlled computations of renormalized quantities while retaining essential features of quantum field theory in curved backgrounds. Moreover, it exhibits distinctive properties such as the absence of a genuine Weyl anomaly in the conformally coupled massless limit, which provides a nontrivial consistency check for renormalization schemes \cite{BirrellDavies1982}.

The main objective of this paper is to compute the renormalized 
energy--momentum tensor for a charged scalar field in $\mathrm{dS}_3$ 
in the presence of a constant background electric field using the 
adiabatic regularization method 
\cite{PhysRevD.9.341,birrell1978application,bunch1980adiabatic,PhysRevD.36.2963}. 
We clarify the relative roles of vacuum polarization and particle production in different physical regimes. In particular, we identify the dominant contributions in the strong-field limit and characterize the infrared behavior associated with light scalar fields.

\subsection{Relation to previous work on scalar QED in de Sitter space}

Quantum electrodynamics in de Sitter spacetime has been widely studied, 
particularly in connection with Schwinger pair production and its interplay 
with cosmological expansion. In lower-dimensional settings such as 
$\mathrm{dS}_2$, it has been shown that an external electric field induces 
particle creation and generates a finite current after proper renormalization 
\cite{PhysRevD.49.6343,PhysRevD.94.104011,PhysRevD.93.025004,hayashinaka2016point}.

In four-dimensional de Sitter spacetime, these analyses reveal a richer structure, 
including nonlinear dependence of the induced current on the electric field and 
infrared enhancement effects for light scalar fields 
\cite{PhysRevD.94.104011,PhysRevD.97.105010,frob2014schwinger,kobayashi2014schwinger}. 
The role of vacuum polarization and its contribution to backreaction has also 
been clarified in this context.

More recently, attention has turned to the renormalized energy--momentum tensor 
as a complementary observable that captures the full energetic content of the 
quantum state. In both two- and four-dimensional settings, finite expressions 
have been obtained using adiabatic regularization and related methods 
\cite{PhysRevD.103.105009,PhysRevD.107.125001}, demonstrating that the 
energy--momentum tensor contains contributions beyond particle production alone.

In contrast, the three-dimensional case remains comparatively unexplored. 
Although $\mathrm{dS}_3$ shares structural features with higher-dimensional 
de Sitter spacetimes, it also exhibits distinctive properties, such as the 
absence of a Weyl anomaly and the topological nature of gravity.
To our knowledge, a renormalized expression for the scalar-field energy-momentum tensor in $\mathrm{dS}_3$, evaluated in the Poincaré patch in the presence of a constant electric background, has not been presented.
\subsection{Main novelty and main claims of this paper}

The main objective of this work is to provide a renormalized derivation of the energy-momentum tensor evaluated in the Poincaré patch, consistent with the appropriate Ward identity for a charged scalar field in three-dimensional de Sitter spacetime in the presence of a constant electric field.
The principal results of the paper can be summarized as follows.
First, we derive an explicit expression for the renormalized energy-momentum tensor 
$\langle T_{\mu\nu} \rangle_{\mathrm{ren}}$ in dS$_3$ using the adiabatic regularization method, 
evaluated in the Poincaré patch and consistent with the appropriate Ward identity for a charged scalar field in an external electromagnetic background.
Second, we identify the leading behavior of the energy density in the strong-field regime, showing that it exhibits a quadratic scaling characteristic of the asymptotic regime, with the dominant contribution arising from vacuum polarization effects.
Third, we analyze the infrared regime associated with light scalar fields and demonstrate the emergence of enhanced contributions with a characteristic inverse dependence on the scalar mass, reflecting the amplification of long-wavelength modes in de Sitter spacetime.
Fourth, we examine the trace of the renormalized energy--momentum tensor and verify that, in the massless conformally coupled limit, the trace vanishes, consistent with the absence of a genuine Weyl anomaly in odd-dimensional spacetimes.
Finally, we perform several internal consistency checks, including the verification of the Ward identity and asymptotic limits. We emphasize that the Bogoliubov analysis presented in this work captures only the particle-production contribution and is used as an auxiliary diagnostic tool; it does not provide an independent reconstruction or equality check for the full renormalized energy--momentum tensor, which also contains state-independent vacuum polarization contributions.

\subsection{Organization of the paper}
The paper is organized as follows. In Sec.~\ref{sec:setup}, we introduce the theoretical setup, including the background geometry, gauge potential, and mode functions. In Sec.~\ref{sec:renorm}, we derive the bare energy--momentum tensor and implement the adiabatic regularization procedure to obtain its renormalized form. In Sec.~\ref{sec:checks}, we present internal consistency checks, including the analysis of the ultraviolet structure and numerical verification. The physical behavior of the renormalized tensor in different regimes is analyzed in Sec.~\ref{sec:behavior}. In Sec.~\ref{sec:trace}, we study the trace of the renormalized energy--momentum tensor in the conformal massless limit. The relation between the renormalized tensor and the Bogoliubov particle-production analysis is discussed in Sec.~\ref{sec:bogoliubov}. Finally, we summarize our results and discuss their implications in Sec.~\ref{sec:conclusion}. The appendices provide supplementary material. 
In Appendix~\ref{app:uv}, we present the details of the ultraviolet numerical check. 
In Appendix~\ref{app:whittaker}, we list the explicit expressions of the momentum integrals involving Whittaker functions.

\section{Setup and mode functions}
\label{sec:setup}
\subsection{Background geometry, gauge potential, and action}
We consider a charged scalar field in the Poincaré patch of $\mathrm{dS}_3$
in the presence of a uniform electric field. The gravitational and
electromagnetic fields are treated as classical backgrounds.
We adopt the metric signature $(+,-,-)$ and use natural units $\hbar=c=1$.
Greek indices $\mu,\nu=0,1,2$ denote spacetime coordinates, while Latin indices $i,j=1,2$ refer to spatial components. The metric in the Poincaré patch can be written as
\begin{align}\label{metric:flrw}
ds^{2}&=dt^{2}-e^{2Ht}d\x^{2}, & t&\in(-\infty,\infty), & \x&\in \mathbb{R},
\end{align}
where $t$ is the proper time and $H$ is the Hubble constant. By using the transformation
\begin{align}\label{conformal}
&\tau=-\frac{1}{H}e^{-Ht}, & \tau \in(-\infty,0)
\end{align}
in a manifestly conformally flat form, the metric as \eqref{metric:flrw} can be expressed
\begin{align}\label{metric:conf}
&ds^{2}=\Omega^{2}(\tau)\big(d\tau^{2}-d\x^{2}\big), & \Omega(\tau)=-\frac{1}{H\tau}.
\end{align}
We choose a gauge potential that generates a uniform electric field, in the metric \eqref{metric:conf}  as
\begin{equation}\label{potential}
A_{\mu}(\tau)=-\frac{E}{H^{2}\tau}\delta_{\mu}^{1},
\end{equation}
where $E$ is a constant.
The action of a charged scalar field $\varphi(x)$ with mass $m$ and charge $e$ is,
\begin{equation}\label{action}
S=\int d^{3}x\sqrt{-g}\Big\{ g^{\mu\nu}\big(\partial_{\mu}+ieA_{\mu}\big)\varphi\big(\partial_{\nu}-ieA_{\nu}\big)\varphi^{\ast}-\big(m^{2}+\xi R\big)\varphi\varphi^{\ast} \Big\},
\end{equation}
where $\xi$ is a dimensionless nonminimal coupling constant and 
 $R=6H^{2}$ represents  the Ricci scalar curvature of $\mathrm{dS}_{3}$, in terms of the Hubble constant $H$.

\subsection{Mode equation and dimensionless parameters}
For the scalar field, we have the equations of motion derived from the Euler-Lagrange  as
\begin{equation}\label{kg}
\frac{1}{\sqrt{-g}}\partial_{\mu}\big(\sqrt{-g}g^{\mu\nu}\partial_{\nu}\varphi\big)+2ieg^{\mu\nu}A_{\mu}\partial_{\nu}\varphi
-e^{2}g^{\mu\nu}A_{\mu}A_{\nu}\varphi+(m^{2}+\xi R)\varphi=0.
\end{equation}
Substituting Eqs.~\eqref{metric:conf} and \eqref{potential} into Eq.~\eqref{kg}, we obtain
\begin{equation}\label{phieq}
\bigg[\rnd_{0}^{2}-\delta^{ij}\rnd_{i}\rnd_{j}+H\Omega(\tau)\rnd_{0}-\frac{2ieE}{H}\Omega(\tau)\rnd_{1}
+\Big(\frac{e^{2}E^{2}}{H^{2}}+(m^{2}+\xi R)\Big)\Omega^{2}(\tau)\bigg]\varphi(x)=0.
\end{equation}
The conformal rescaling of the scalar field 
\begin{equation}\label{varphi}
\tilde{\varphi}(x):=\Omega^{\frac{1}{2}}(\tau)\varphi(x),
\end{equation}
 yields the following Klein–Gordon equation
\begin{equation}\label{vphieq}
\bigg[\rnd_{0}^{2}-\delta^{ij}\rnd_{i}\rnd_{j}+\frac{2ieE}{\tau H^{2}}\rnd_{1}+\frac{1}{\tau^{2}}
\Big(\frac{e^{2}E^{2}}{H^{4}}+\frac{(m^{2}+\xi R)}{H^{2}}
-\frac{3}{4}\Big)\bigg]\tilde{\varphi}(x)=0.
\end{equation}
The invariance of Eq.~\eqref{vphieq} under translation along the spatial directions is defined as
\begin{equation}\label{fpm}
\tilde{\varphi}(\tau,\mathrm{x})=e^{\pm i\mathrm{k}\cdot\x}f^{\pm}(\tau),
\end{equation}
In this case, the superscript $\pm$ denotes the positive and negative frequency solutions, respectively.
We define $k = |\mathbf{k}|$ and introduce the dimensionless ratio
\begin{equation}
r = \frac{k_x}{k},
\end{equation}
The quantity $r$ measures the projection of the momentum along the electric-field direction.
Substituting Eq.~\eqref{fpm} into Eq.~\eqref{vphieq} leads to
\begin{equation}\label{feq}
\frac{d^{2}}{dz_{\pm}^{2}}f^{\pm}(z_{\pm})+\Big(-\frac{1}{4}+\frac{\kappa}{z_{\pm}}
+\frac{1/4-\gamma^{2}}{z_{\pm}^{2}}\Big)f^{\pm}(z_{\pm})=0,
\end{equation}
where the variables $z_{+}$ and $z_{-}$ are defined as follows
\begin{align}\label{zpm}
z_{+}&:=+2ik\tau, & z_{-}&:=e^{i\pi}z_{+}=-2ik\tau,
\end{align}.
The dimensionless parameters are defined by
\begin{align}\label{paramete}
\lambda_{\mathrm{m}} &= \frac{m}{H}, 
& \lambda &= -\frac{eE}{H^{2}}, \\
\bar{\xi} &= \xi - \frac{1}{8}, 
& \kappa &= -i\lambda r, \\
\gamma &= \sqrt{\frac{1}{4} - \lambda^{2} - \lambda_{\mathrm{m}}^{2} - 6\bar{\xi}}.
\end{align}
The parameters $\lambda_m$ and $\lambda$ characterize the mass of the scalar field and the strength of the electric field relative to the Hubble scale, respectively. The parameter $\kappa$ encodes the coupling between the momentum and the electric field, while $\gamma$ determines the effective index of the mode functions and controls their infrared behavior and mass dependence.

\subsection{In/out modes and vacuum choice}
In order to define the vacuum state and the associated particle concept, we construct a complete set of mode functions that specify the annihilation and creation operators. These modes are selected by imposing appropriate asymptotic conditions at early and late times \cite{BirrellDavies1982, kim2016schwinger}.
At early times $\tau \to -\infty$, the effective frequency approaches a constant and the positive- and negative-frequency solutions are uniquely identified by their Minkowski-like behavior. Using the asymptotic properties of the Whittaker functions, the normalized in-modes can be written as
\begin{align}
\label{uinb}
U_{in\mathbf{k}}(x)&=(2k)^{-\frac{1}{2}}e^{\frac{i\pi\kappa}{2}}\Omega^{-\frac{1}{2}}(\tau)e^{+i\mathbf{k}\cdot\mathbf{x}} W_{\kappa,\gamma}(z_{+}), \\
\label{vinb}
V_{in\mathbf{k}}(x)&=(2k)^{-\frac{1}{2}}e^{-\frac{i\pi\kappa}{2}}\Omega^{-\frac{1}{2}}(\tau)e^{-i\mathbf{k}\cdot\mathbf{x}} W_{\kappa,-\gamma}(z_{-}),
\end{align}
which behave as positive- and negative-frequency modes in the remote past.
When the parameters $\kappa,~\gamma$, and the phase of the variable $z$, satisfy these conditions
\begin{align}\label{condit}
\frac{1}{2}\pm\gamma-\kappa \neq 0,-1,-2,\ldots, && \big|\mathrm{ph}(z)\big|<\frac{3}{2}\pi,
\end{align}
the Whittaker function $W_{\kappa,\gamma}(z)$ has the Mellin–Barnes integral representation
\begin{equation}\label{Mellin}
W_{\kappa,\gamma}(z)=e^{-\frac{z}{2}}\int_{-i\infty}^{+i\infty}\frac{ds}{2\pi i}
\frac{\Gamma\big(\frac{1}{2}+\gamma+s\big) \Gamma\big(\frac{1}{2}-\gamma+s\big)\Gamma\big(-\kappa-s\big)}
{\Gamma\big(\frac{1}{2}+\gamma-\kappa\big)\Gamma\big(\frac{1}{2}-\gamma-\kappa\big)}z^{-s},
\end{equation}
where $\Gamma(z)$ denotes the Gamma function and the contour of integration separates the poles of 
$\Gamma(1/2+\gamma+s)\Gamma(1/2-\gamma+s)$ from those of $\Gamma(-\kappa-s)$.
Similarly, in the asymptotic future $\tau \to 0^{-}$, the dominant contribution to the frequency arises from the $1/\tau^2$ terms, allowing one to define a corresponding set of out-modes. The normalized solutions in this regime take the form
\begin{align}
\label{uoutb}
U_{out\mathbf{k}}(x)&=(4|\gamma|k)^{-\frac{1}{2}}e^{\frac{i\pi\gamma}{2}}\Omega^{-\frac{1}{2}}(\tau)e^{+i\mathbf{k}\cdot\x}{\mathrm{M}_{\kappa,\gamma}}(z_{+}),
\\ \label{voutb}
V_{out\mathbf{k}}(x)&=(4|\gamma|k)^{-\frac{1}{2}}e^{\frac{i\pi\gamma}{2}}\Omega^{-\frac{1}{2}}(\tau)e^{-i\mathbf{k}\cdot\x}{\mathrm{M}_{\kappa,-\gamma}}(z_{-}).
\end{align}
The labels ``in'' and ``out'' refer to the fact that these modes reduce to positive-frequency solutions at early and late times, respectively, and define the corresponding in- and out-vacua.
The normalization of the mode functions is fixed by the conserved Klein--Gordon inner product,
\begin{equation}
(u_1,u_2)= i \int d^{2}x \sqrt{-g}\, g^{0\nu} \left(u_1^* \partial_\nu u_2 - u_2 \partial_\nu u_1^* \right),
\end{equation}
evaluated on a constant-time hypersurface. With this choice, the in- and out-modes satisfy the standard orthonormality relations, ensuring the consistency of the quantization procedure.
The orthonormality of the mode functions is ensured by the existence of a time-independent conserved inner product. For two solutions $u_1(x)$ and $u_2(x)$ of the field equation, this product is defined on a constant-time hypersurface. With respect to this inner product, the in- and out-modes form
complete orthonormal sets, satisfying the standard relations
\begin{align}\label{orthonormb}
\big(U_{in(out)\mathbf{k}},U_{in(out)\mathbf{k}'}\big)&=-\big(V_{in(out)\mathbf{k}},V_{in(out)\mathbf{k}'}\big)
=(2\pi)^{2}\delta^{(2)}(\mathbf{k}-\mathbf{k}'), \nn\\
\big(U_{in(out)\mathbf{k}},V_{in(out)\mathbf{k}'}\big)&=0.
\end{align}
We constructed two complete sets of orthonormal mode functions, 
$\{U_{\text{in},\mathbf{k}},V_{\text{in},\mathbf{k}}\}$  and 
$\{U_{\text{out},\mathbf{k}},V_{\text{out},\mathbf{k}}\}$, corresponding to early-
and late-time particle definitions, respectively. In this section, we use these
modes to describe particle creation through a Bogoliubov transformation.
The scalar field operator can be expanded in terms of the in-modes as
\begin{equation}
\phi(x)=\int\frac{d^2 k}{(2\pi)^2}
\left[U_{\text{in},\mathbf{k}}(x)a_{\text{in},\mathbf{k}}
+V_{\text{in},\mathbf{k}}(x)b_{\text{in},\mathbf{k}}^\dagger\right],
\label{phi-in-expansionb}
\end{equation}
where the operators satisfy the canonical commutation relations
\begin{equation}
[a_{\text{in},\mathbf{k}},a_{\text{in},\mathbf{k}'}^\dagger]
=[b_{\text{in},\mathbf{k}},b_{\text{in},\mathbf{k}'}^\dagger]
=(2\pi)^2\delta^{(2)}(\mathbf{k}-\mathbf{k}'),
\label{in-commb}
\end{equation}
and define the in-vacuum by
\begin{equation}
a_{\text{in},\mathbf{k}}|0\rangle_{\text{in}}=0.
\label{in-vacuumb}
\end{equation}
However, since there is no global timelike Killing vector, a unique definition of positive-frequency modes is not available, and therefore the scalar field admits an alternative expansion in terms of another complete set of orthonormal modes.
Similarly, the field can be expanded in the out-basis as
\begin{equation}
\phi(x)=\int\frac{d^2 k}{(2\pi)^2}
\left[U_{\text{out},\mathbf{k}}(x)a_{\text{out},\mathbf{k}}
+V_{\text{out},\mathbf{k}}(x)b_{\text{out},\mathbf{k}}^\dagger\right].
\label{phi-out-expansionb}
\end{equation}
In this case, the commutation relations are
\begin{equation}
\left[a_{\text{out}\,\mathbf{k}},\, a^{\dagger}_{\text{out}\,\mathbf{k}'}\right]
=
\left[b_{\text{out}\,\mathbf{k}},\, b^{\dagger}_{\text{out}\,\mathbf{k}'}\right]
=
(2\pi)^2 \delta^{(2)}(\mathbf{k}-\mathbf{k}'),
\end{equation}
with a new vacuum state
\begin{equation}
a_{\text{out},\mathbf{k}}|0\rangle_{\text{out}}=0.
\label{out-vacuumb}
\end{equation}
The mode functions \eqref{uinb} and \eqref{vinb} fulfill the Wronskian condition
\begin{equation}\label{wronskian}
U_{in,\mathbf{k}}\dot{U}_{in,\mathbf{k}}^{\ast}-U_{in,\mathbf{k}}^{\ast}\dot{U}_{in,\mathbf{k}}=V_{in\mathbf{k}}^{\ast}\dot{V}_{in,\mathbf{k}}-V_{in,\mathbf{k}}\dot{V}_{in,\mathbf{k}}^{\ast}=i\Omega^{-1}(\tau).
\end{equation}

\section{Bare energy-momentum tensor and adiabatic renormalization}
\label{sec:renorm}

\subsection{In-vacuum expectation values of bare energy-momentum tensor}
The energy-momentum tensor of the scalar field is
defined by variation of the action $\delta S$ with respect to the inverse metric $\delta g^{\mu\nu}$ as
\begin{equation}\label{actiongeneral}
T_{\mu\nu}=+\frac{2}{\sqrt{-g}}\frac{\delta S}{\delta g^{\mu\nu}}.
\end{equation}
 From a straightforward calculation \eqref{actiongeneral}, we obtain the symmetric expression for the energy–momentum tensor of the scalar field as
\begin{eqnarray}\label{emt:sc}
T_{\mu\nu} &=& \Big[\big(4\xi-1\big)g^{\rho\sigma}\big(\partial_{\rho}+ieA_{\rho}\big)\varphi\big(\partial_{\sigma}-ieA_{\sigma}\big)\varphi^{\ast}
+\big(1-4\xi\big)m^{2}\varphi\varphi^{\ast}+\Big(\frac{1}{3}-4\xi\Big)\xi R\varphi\varphi^{\ast}\Big]g_{\mu\nu} \nn\\
&+&\big(1-2\xi\big)\Big(\partial_{\mu}\varphi\partial_{\nu}\varphi^{\ast}+\partial_{\nu}\varphi\partial_{\mu}\varphi^{\ast}\Big)
+ieA_{\mu}\Big(\varphi\partial_{\nu}\varphi^{\ast}-\varphi^{\ast}\partial_{\nu}\varphi\Big)
+ieA_{\nu}\Big(\varphi\partial_{\mu}\varphi^{\ast}-\varphi^{\ast}\partial_{\mu}\varphi\Big) \nn\\
&+&2e^{2}A_{\mu}A_{\nu}\varphi\varphi^{\ast}+2\xi\Gamma^{\rho}_{\mu\nu}\Big(\varphi\partial_{\rho}\varphi^{\ast}+\varphi^{\ast}\partial_{\rho}\varphi\Big)
-2\xi\Big(\varphi\partial_{\mu}\partial_{\nu}\varphi^{\ast}+\varphi^{\ast}\partial_{\mu}\partial_{\nu}\varphi\Big).
\end{eqnarray}
The nonzero Christoffel symbols for the metric \eqref{metric:conf} are as follows
\begin{align}\label{christoff}
\Gamma^{0}_{00}&=\frac{\dot{\Omega}}{\Omega}, & \Gamma^{0}_{ij}&= \frac{\dot{\Omega}}{\Omega}\delta_{ij}, &
\Gamma^{i}_{0j}&= \frac{\dot{\Omega}}{\Omega}\delta^{i}_{j},
\end{align}
where the indices $i,j$ denote only two spatial components. By using Eq.~\eqref{christoff}, the Ricci tensor and hence the Ricci scalar can be calculated
\begin{align}\label{ricci}
R_{\mu\nu}&=2H^{2}g_{\mu\nu}, & R&=6H^{2}.
\end{align}
Integral representations for the in-vacuum expectation values of the components of the energy-momentum tensor can be obtained by substituting the mode  expansion~\eqref{phi-in-expansionb} for the quantum scalar field $\varphi(x)$ into the definition \eqref{emt:sc}.  Using equations of motion~\eqref{kg} and some algebraic manipulations we obtain the expectation values of the energy-momentum tensor components. The integral expression of the timelike component is given by
\begin{align}\label{vint00}
&\big\langle\mathrm{in}\big|T_{00}\big|\mathrm{in}\big\rangle = \int\frac{d^{2}k}{(2\pi)^{2}}\bigg[\dot{U}_{\mathrm{k}}\dot{U}_{\mathrm{k}}^{\ast}-4\xi\tau^{-1}\Big(U_{\mathrm{k}}\dot{U}_{\mathrm{k}}^{\ast}+\dot{U}_{\mathrm{k}}U_{\mathrm{k}}^{\ast}\Big)
+\tau^{-2}\Big(k^{2}\tau^{2}+2\lambda rk\tau+\lambda^{2}+\lambda_{m}^{2}
\nn\\
&+2\xi\Big) U_{\mathrm{k}}U_{\mathrm{k}}^{\ast}\bigg].
\end{align}
For the diagonal spacelike components we get
\begin{align}\label{vint11}
&\big\langle\mathrm{in}\big|T_{11}\big|\mathrm{in}\big\rangle = \int\frac{d^{2}k}{(2\pi)^{2}}\bigg\{\big(1-4\xi\big)\dot{U}_{\mathrm{k}}\dot{U}_{\mathrm{k}}^{\ast}
-2\xi\tau^{-1}\Big(U_{\mathrm{k}}\dot{U}_{\mathrm{k}}^{\ast}+\dot{U}_{\mathrm{k}}U_{\mathrm{k}}^{\ast}\Big)
+\tau^{-2}\Big[\big(4\xi-1+2r^{2}\big)k^{2}\tau^{2} \nn\\
&+2\big(4\xi+1\big)\lambda kr\tau+\big(4\xi+1\big)\lambda^{2}+\big(4\xi-1\big)\lambda_{m}^{2}+2\xi\big(12\xi-1\big)\Big]U_{\mathrm{k}}U_{\mathrm{k}}^{\ast}\bigg\},
\end{align}
and
\begin{align}\label{vint22}
&\big\langle\mathrm{in}\big|T_{22}\big|\mathrm{in}\big\rangle =
 \int\frac{d^{2}k}{(2\pi)^{2}}\bigg\{\big(1-4\xi\big)\dot{U}_{\mathbf{k}}\dot{U}_{\mathbf{k}}^{\ast}
-2\xi\tau^{-1}\Big(U_{\mathbf{k}}\dot{U}_{\mathbf{k}}^{\ast}+\dot{U}_{\mathbf{k}}U_{\mathbf{k}}^{\ast}\Big)
+\tau^{-2}\Big[\big(4\xi-1\big) k^{2}\tau^{2}\nn\\
&+2k_{y}^{2}\tau^{2}+2\big(4\xi-1\big)\lambda kr\tau+\big(4\xi-1\big)\lambda^{2}+\big(4\xi-1\big)\lambda_{m}^{2}
+2\xi\big(12\xi-1\big)\Big]
 U_{\mathbf{k}}U_{\mathbf{k}}^{\ast}\bigg\}.
\end{align}
The only nonvanishing in-vacuum expectation values of the off-diagonal components can be expressed as
\begin{equation}\label{vint01}
\big\langle\mathrm{in}\big|T_{01}\big|\mathrm{in}\big\rangle = \big\langle\mathrm{in}\big|T_{10}\big|\mathrm{in}\big\rangle
= i\tau^{-1}\int\frac{d^{2}k}{(2\pi)^{2}}\big(rk\tau+\lambda\big)\Big(U_{\mathbf{k}}\dot{U}_{\mathbf{k}}^{\ast}-\dot{U}_{\mathbf{k}}U_{\mathbf{k}}^{\ast}\Big).
\end{equation}
Changing the integral variable from the comoving  momentum $k$, to
the dimensionless physical momentum $p=-k\tau$, and imposing an ultraviolet cutoff $\Lambda$ on $p$,
the in-vacuum expectation value of the timelike component \eqref{vint00} can be expressed as
\begin{align}\label{timeli}
&\big \langle \mathrm{in} \big| T_{00} \big| \mathrm{in} \big \rangle = \Omega^{2}(\tau)\frac{H^{3}}{(2\pi)^{2}}\int_{-1}^{1}\frac{dr}{\sqrt{1-r^{2}}}
\bigg[2\mathcal{I}_{1}-4\lambda r\mathcal{I}_{2}+\Big(\frac{1}{4}-\gamma^{2}-8\bar{\xi}+\lambda^{2}r^{2}\Big)\mathcal{I}_{3}
+\mathcal{I}_{4} \nn\\
&-\lambda r\mathcal{I}_{5}+(4\xi-\frac{1}{2})\mathcal{I}_{6} +\mathcal{I}_{7} \bigg],
\end{align}
 where $\mathcal{I}_{1},\,\mathcal{I}_{2},\ldots,\mathcal{I}_{7}$ defined in  Eqs.~(\ref{def:I1})-(\ref{def:I7}) as the momentum integrals over the Whittaker functions. The in-vacuum expectation value of the spacelike components \eqref{vint11}  and \eqref{vint22} are expressed by
\begin{align}\label{spaceli1}
&\big \langle \mathrm{in} \big| T_{11} \big| \mathrm{in} \big \rangle=\Omega^{2}(\tau)\frac{H^{3}}{(2\pi)^{2}}\int_{-1}^{1}\frac{dr}{\sqrt{1-r^{2}}}
\bigg[2r^{2} \mathcal{I}_{1}-4\lambda r\mathcal{I}_{2} 
+\big\{ (4\xi+1)\lambda^{2}+(4\xi-1)\lambda_{m}^{2} \nn\\
&-(4\xi-1)\lambda^{2} r^{2}+2\xi(12\xi-2)+\frac{1}{4}\big\}\mathcal{I}_{3}
+\big(1-4\xi\big)\mathcal{I}_{4} 
-\big(1-4\xi\big) \lambda r \mathcal{I}_{5}+\big(4\xi-\frac{1}{2}\big)\mathcal{I}_{6} \nn\\
&- \big(4\xi-1\big)\mathcal{I}_{7} \bigg],
\end{align}
and
\begin{align}\label{spaceli2}
&\big \langle \mathrm{in} \big| T_{22} \big| \mathrm{in} \big \rangle=\Omega^{2}(\tau)\frac{H^{3}}{(2\pi)^{2}}\int_{-1}^{1}\frac{dr}{\sqrt{1-r^{2}}}
\bigg[2(1-r^{2}) \mathcal{I}_{1} 
+\big\{ (4\xi-1)(\lambda^{2}+\lambda_{m}^{2}-\lambda^{2} r^{2})  \nn\\
&+2\xi(12\xi-2)+\frac{1}{4}\big\}\mathcal{I}_{3}
+\big(1-4\xi\big)\mathcal{I}_{4} 
-\big(1-4\xi\big) \lambda r \mathcal{I}_{5}+\big(4\xi-\frac{1}{2}\big)\mathcal{I}_{6} 
-\big(4\xi-1\big)\mathcal{I}_{7} \bigg].
\end{align}
By using Wronskian~\eqref{wronskian}, the in-vacuum expectation values of the off-diagonal components  equal to
\begin{equation}\label{offdiagg}
\big\langle \mathrm{in} \big| T_{01} \big|\mathrm{in} \big\rangle = \big\langle \mathrm{in} \big| T_{10} \big| \mathrm{in} \big\rangle
=\Omega^{2}(\tau)\frac{H^{3}}{(2\pi)^{2}} \big(\pi \lambda\Lambda^{2}\big).
\end{equation}
Substituting the expressions \eqref{endi7}-\eqref{endi6} into Eqs.~\eqref{timeli}, \eqref{spaceli1} and \eqref{spaceli2}, gives the unregularized in-vacuum expectation values of
the timelike and spacelike components of the energy-momentum tensor, the unregularized timelike component is given by
\begin{flalign}\label{vinemt00}
&\big\langle\mathrm{in}\big|T_{00}\big|\mathrm{in}\big\rangle = \Omega^{2}(\tau)\frac{H^{3}}{4\pi^{2}}\bigg\{
\frac{2}{3}\pi\Lambda^{3}
+\Big(\frac{1}{4}-2\xi+\frac{1}{2}\lambda^{2}+\lambda_{m}^{2}\Big)\pi\Lambda
-\frac{3}{32}i \pi+\frac{5}{12}i \pi \gamma^{2}-\frac{1}{2}\pi^{2} \gamma^{2} \nn\\
&-\frac{1}{6} i\pi \gamma^{4}+\frac{1}{2} \pi^{2} \gamma^{4}-\frac{7}{24} i \pi \lambda^{2}-\frac{1}{2} i\pi \gamma^{2}\lambda^{2}-\frac{5}{16} i \pi\lambda^{4} + \Big(-\frac{13}{24}+\frac{1}{6}\gamma^{2}+4\xi+\frac{1}{4}\lambda^{2}\Big)\nn\\
&\times\pi \gamma\cot\big(2\pi\gamma\big)+\Big(\frac{2}{3}-\frac{2}{3}\gamma^{2}+4\xi-\lambda^{2}\Big) \pi\gamma \csc\big(2\pi\gamma\big){\mathrm{I}}_{0}\big(2\pi\lambda\big)+\frac{1}{2}\gamma\lambda \csc\big(2\pi\gamma\big) {\mathrm{I}}_{1}\big(2\pi\lambda\big) \nn\\
&+i\csc\big(2\pi\gamma\big)\int_{-1}^{+1}\frac{dr}{\sqrt{1-r^{2}}}B_{0r}\nn\\
&\bigg[\Big(e^{2\pi\lambda r}+e^{-2i\pi\gamma}\Big)
\psi\Big(\frac{1}{2}-\gamma+i\lambda r\Big)
-\Big(e^{2\pi\lambda r}+e^{2i\pi\gamma}\Big)\psi\Big(\frac{1}{2}+\gamma+i\lambda r\Big)\bigg] ,&&
\end{flalign}
where $\mathrm{I}_{0}$ and $\mathrm{I}_{1}$ are the first and second kind of the Bessel functions respectively, and $\psi$ denotes the digamma function which is given by the logarithmic
derivative of the gamma function.
 The coefficient $B_{0r}$ is given by
\begin{equation}\label{c0}
B_{0r}=\frac{1}{2}\lambda^{3}r^{3}+\frac{1}{8}\Big(16\xi+4\gamma^{2}-1\Big)\lambda r.
\end{equation}
Also, we find the unregularized in-vacuum expectation values of the diagonal spacelike components are
\begin{flalign}\label{vinemt11}
&\big\langle\mathrm{in}\big|T_{11}\big|\mathrm{in}\big\rangle = \Omega^{2}(\tau)\frac{H^{3}}{4\pi^{2}}\bigg\{
\frac{\pi}{3}\Lambda^{3}
+\Big(-\frac{1}{8}+\xi+\frac{5}{8}\lambda^{2}-\frac{1}{2}\lambda_{m}^{2}\Big)\pi\Lambda
-\frac{3}{32} i\pi+\frac{5}{12}i \pi \gamma^{2}-\frac{1}{2}\pi^{2} \gamma^{2}\nn\\
&-\frac{1}{6} i\pi \gamma^{4} +\frac{1}{2} \pi^{2} \gamma^{4}+\frac{3}{8}i\pi \xi-\frac{5}{3}i\pi\xi \gamma^{2}+ 2\pi^{2}\xi\gamma^{2}+\frac{2}{3} i\pi\gamma^{4}\xi-2\pi^{2}\gamma^{4}\xi-\frac{7}{24} i\pi\lambda^{2}-\frac{1}{2} i\pi\lambda^{2}\gamma^{2}  \nn\\
&+\frac{7}{6} i\pi\xi\lambda^{2}+2 i\pi\xi\lambda^{2}\gamma^{2}-\frac{5}{16}i\pi\lambda^{4}+\frac{5}{4}i\pi\xi\lambda^{4}+\Big(-\frac{5}{24}-\frac{1}{6}\gamma^{2}+\frac{9}{2}\xi-2\gamma^{2}\xi-24\xi^{2}-\frac{1}{8}\lambda^{2}
\nn\\
&-3 \xi \lambda^{2}+\lambda_{m}^{2}-4 \xi \lambda_{m}^{2}\Big) \pi \gamma\cot\big(2\pi\gamma\big)+\Big(-\frac{15}{4}\frac{1}{\pi^{2}}+\frac{5}{3}-\frac{5}{3}\gamma^{2}-\frac{1}{3}\xi+\frac{4}{3}\xi\gamma^{2}-24\xi^{2}-2\lambda^{2} \nn\\
&+4\xi\lambda^{2}+\lambda_{m}^{2}-4\xi \lambda_{m}^{2}\Big) \pi\gamma\csc\big(2\pi\gamma\big){\mathrm{I}}_{0}\big(2\pi\lambda\big)+\Big(-\frac{2}{3\lambda}+\frac{15}{4}\frac{1}{\pi^{2}\lambda}+\frac{2}{3}\frac{\gamma^{2}}{\lambda}+3\lambda-4\xi \lambda\Big) \nn\\
&\times \gamma \csc\big(2\pi\gamma\big) {\mathrm{I}}_{1}\big(2\pi\lambda\big) +i\csc\big(2\pi\gamma\big)\int_{-1}^{+1}\frac{dr}{\sqrt{1-r^{2}}}B_{1r} \nn\\
&\bigg[\Big(e^{2\pi\lambda r}+e^{-2i\pi\gamma}\Big)
\psi\Big(\frac{1}{2}-\gamma+i\lambda r\Big)
-\Big(e^{2\pi\lambda r}+e^{2i\pi\gamma}\Big)\psi\Big(\frac{1}{2}+\gamma+i\lambda r\Big)\bigg] ,&&
\end{flalign}
where the coefficient $B_{1r}$ is given by
\begin{eqnarray}\label{b1}
B_{1r}&=&-\frac{5}{2}r^{5}\lambda^{3}+\Big(\frac{7}{8}\lambda-\frac{3}{2}\lambda\gamma^{2}+3\lambda^{3}\Big)r^{3}
+\Big(-\frac{3}{4}\lambda+\frac{3}{2}\gamma^{2}\lambda+\frac{5}{2}\xi\lambda-2\xi\gamma^{2}\lambda-12\xi^{2}\lambda \nn\\
&-&\frac{1}{2}\lambda^{3}-2\xi\lambda^{3}+\frac{1}{2}\lambda \lambda_{m}^{2}-2\xi
\lambda \lambda_{m}^{2}\Big)r,
\end{eqnarray}
we also have
\begin{flalign}\label{vinemt22}
&\big\langle\mathrm{in}\big|T_{22}\big|\mathrm{in}\big\rangle = \Omega^{2}(\tau)\frac{H^{3}}{4\pi^{2}}\bigg\{
\frac{\pi}{3}\Lambda^{3}
+\Big(-\frac{1}{8}+\xi-\frac{1}{8}\lambda^{2}-\frac{1}{2}\lambda_{m}^{2}\Big)\pi\Lambda
-\frac{3}{32} i\pi+\frac{5}{12}i \pi \gamma^{2} \nn\\
&-\frac{1}{2}\pi^{2} \gamma^{2}-\frac{1}{6} i\pi \gamma^{4}+\frac{1}{2} \pi^{2} \gamma^{4}+\frac{3}{8}i\pi \xi-\frac{5}{3}i\pi\xi \gamma^{2}+ 2\pi^{2}\xi\gamma^{2}+\frac{2}{3} i\pi\gamma^{4}\xi-2\pi^{2}\gamma^{4}\xi-\frac{7}{24} i\pi\lambda^{2} \nn\\
&-\frac{1}{2} i\pi\lambda^{2}\gamma^{2} +\frac{7}{6} i\pi\xi\lambda^{2}+2 i\pi\xi\lambda^{2}\gamma^{2}-\frac{5}{16}i\pi\lambda^{4}+\frac{5}{4}i\pi\xi\lambda^{4}+\Big(-\frac{5}{24}-\frac{1}{6}\gamma^{2}+\frac{9}{2}\xi-2\gamma^{2}\xi
\nn\\
&-24\xi^{2}+\frac{1}{8}\lambda^{2}-3 \xi \lambda^{2}+\lambda_{m}^{2}-4 \xi \lambda_{m}^{2}\Big) \pi \gamma\cot\big(2\pi\gamma\big)+\Big(\frac{15}{4}\frac{1}{\pi^{2}}+\frac{1}{3}-\frac{1}{3}\gamma^{2}-\frac{1}{3}\xi+\frac{4}{3}\xi\gamma^{2}\nn\\
&-24\xi^{2}-\lambda^{2}+4\xi\lambda^{2}+\lambda_{m}^{2}-4\xi \lambda_{m}^{2}\Big) \pi\gamma \csc\big(2\pi\gamma\big){\mathrm{I}}_{0}\big(2\pi\lambda\big)+\Big(\frac{2}{3\lambda}-\frac{15}{4}\frac{1}{\pi^{2}\lambda}-\frac{2}{3}\frac{\gamma^{2}}{\lambda}-\frac{3}{2}\lambda \nn\\
&-4\xi \lambda\Big)\gamma \csc\big(2\pi\gamma\big){\mathrm{I}}_{1}\big(2\pi\lambda\big) +i\csc\big(2\pi\gamma\big)\int_{-1}^{+1}\frac{dr}{\sqrt{1-r^{2}}}B_{2r}\nn\\
&\bigg[\Big(e^{2\pi\lambda r}+e^{-2i\pi\gamma}\Big)
\psi\Big(\frac{1}{2}-\gamma+i\lambda r\Big)
-\Big(e^{2\pi\lambda r}+e^{2i\pi\gamma}\Big)\psi\Big(\frac{1}{2}+\gamma+i\lambda r\Big)\bigg] ,&&
\end{flalign}
where the coefficient $B_{2r}$ is given by
\begin{eqnarray}\label{b2}
B_{2r}&=&\frac{5}{2}r^{5}\lambda^{3}+\Big(-\frac{7}{8}\lambda+\frac{3}{2}\lambda\gamma^{2}-\frac{5}{2}\lambda^{3}\Big)r^{3}
+\Big(\frac{3}{8}\lambda-\gamma^{2}\lambda+\frac{5}{2}\xi\lambda-2\xi\gamma^{2}\lambda-12\xi^{2}\lambda \nn\\
&+&\frac{1}{2}\lambda^{3}-2\xi\lambda^{3}+\frac{1}{2}\lambda \lambda_{m}^{2}-2\xi
\lambda \lambda_{m}^{2}\Big)r,
\end{eqnarray}

\subsection{Adiabatic expansion and counterterms} \label{Adia:expansion}
We employ the adiabatic regularization procedure to eliminate the divergent terms of the expressions \eqref{vinemt00}, \eqref{vinemt11} and \eqref{vinemt22}.  The adiabatic regularization method subtracts the appropriate adiabatic counterterms from the corresponding unregularized expressions. We assume that the electromagnetic vector potential is of adiabatic order zero. We consider the solution to the Klein-Gordon Eq.~\eqref{vphieq} by investigating its positive frequency, as follows:
\begin{equation}\label{adimode}
f(\tau,\x)=e^{+ik\x}\mathcal{U}_{A}(\tau).
\end{equation}
Then the function $\mathcal{U}_{A}(\tau)$ satisfies the following field equation
\begin{equation}\label{klein}
\frac{d^{2}\mathcal{U}_{A}}{d\tau^{2}}+\Big(\omega_{0}^{2}(\tau)+\Delta(\tau)\Big)\mathcal{U}_{A}=0,
\end{equation}
where $\mathcal{U}_{A}$  is the adiabatic solution with a positive frequency, and the conformal time dependent frequencies are given by
\begin{eqnarray}
\omega_{0}(\tau) &=& \Big(k^{2}+2eA_{1}kr+e^{2}A_{1}^{2}+m^{2}\Omega^{2}\Big)^{\frac{1}{2}}, \label{omega}\\
\Delta(\tau) &=& 6\xi\Big(\frac{\dot{\Omega}}{\Omega}\Big)^{2}, \label{delta}
\end{eqnarray}
in this case, $\omega_{0}$ is zero adiabatic order while $\Delta$ is second adiabatic order. For our purpose, we expand our modes up to second adiabatic order and subtract from the corresponding original expressions for the in-vacuum expectation values  leads to the regularized induced energy-momentum tensor. We begin by considering the Wentzel-Kramers-Brillouin (WKB) type solution for Eq.~\eqref{klein} as 
\begin{equation}\label{wkb}
\mathcal{U}_{A}(\tau)=\frac{1}{\sqrt{2\mathcal{W}(\tau)}}\exp\Big[-i\int^{\tau}\mathcal{W}(\tau')d\tau'\Big],
\end{equation}
where  $\mathcal{W}$ corresponds to the equation
\begin{equation}\label{exact}
\mathcal{W}^{2}=\omega_{0}^{2}+\Delta-\frac{\ddot{\mathcal{W}}}{2\mathcal{W}}+\frac{3\dot{\mathcal{W}}^{2}}{4\mathcal{W}^{2}}.
\end{equation}
In order to put the solution in the desired form, it is convenient to write $\mathcal{W}$ as follows
\begin{equation}\label{mathcalw}
\mathcal{W}=\mathcal{W}^{(0)}+\mathcal{W}^{(2)},
\end{equation}
By substituting the expansion \eqref{mathcalw} into the expression Eq.~\eqref{exact}, we obtain the approximations for the zero and second adiabatic order
\begin{eqnarray}
\mathcal{W}^{(0)}&=&\omega_{0}, \\ \label{w0}
\mathcal{W}^{(2)}&=&\frac{\Delta}{2\omega_{0}}-\frac{\ddot{\omega}_{0}}{4\omega_{0}^{2}}+\frac{3\dot{\omega}_{0}^{2}}{8\omega_{0}^{3}} \label{w2}.
\end{eqnarray}
The adiabatic expansion of $\mathcal{W}(\tau)$ up to second order is obtained from Eqs.~\eqref{mathcalw},\eqref{w0} and \eqref{w2} as follows
\begin{equation}\label{wuptwo}
\mathcal{W}(\tau)=\omega_{0}(\tau)
+\frac{1}{2\omega_{0}}\bigg(\Delta-\frac{\ddot{\omega}_{0}}{2\omega_{0}}+\frac{3\dot{\omega}_{0}^{2}}{4\omega_{0}^{2}}\bigg).
\end{equation}
In addition, we need the adiabatic expansion of $\mathcal{W}^{-1}(\tau)$, which up to second order is given by
\begin{equation}\label{winv}
\frac{1}{\mathcal{W}(\tau)}=\frac{1}{\omega_{0}(\tau)}
-\frac{1}{2\omega_{0}^{3}}\bigg(\Delta-\frac{\ddot{\omega}_{0}}{2\omega_{0}}+\frac{3\dot{\omega}_{0}^{2}}{4\omega_{0}^{2}}\bigg).
\end{equation}
The adiabatic expansion of positive frequency mode function up to second order can be determined by putting together the equations of the (\ref{wkb}), (\ref{wuptwo}), and (\ref{winv}) of Eq.~(\ref{adimode}).
In this case, the counterterms are obtained by taking the adiabatic expansion of the scalar field operator into Eqs.~\eqref{vint00}-~\eqref{vint22} and computing the expectation values of the resulting expressions in the adiabatic vacuum. we find the counterterm to second adiabatic order  for the timelike component
\begin{eqnarray}\label{delta00}
\mathcal{T}_{00}^{(2)} &=& \Omega^{2}(\tau)\frac{H^{3}}{4\pi^{2}}\bigg[
\frac{2}{3}\pi\Lambda^{3}+\Big(\frac{1}{4}-2\xi+\frac{1}{2}\lambda^{2}+\lambda_{m}^{2}\Big)\pi\Lambda+\frac{\pi}{12}\frac{\lambda^{2}}{\lambda_{m}}+\frac{\pi}{3}\lambda_{m}-2\pi\xi\lambda_{m} \nn\\
&-&\frac{2\pi}{3}\lambda_{m}^{3} \bigg].
\end{eqnarray}
We find the counterterms to second order  adiabatic  for the diagonal spacelike components
\begin{eqnarray}\label{delta11}
\mathcal{T}_{11}^{(2)} &=& \Omega^{2}(\tau)\frac{H^{3}}{4\pi^{2}}\bigg[
\frac{\pi}{3}\Lambda^{3}+\Big(-\frac{1}{8}+\xi+\frac{5}{8}\lambda^{2}-\frac{\lambda_{m}^{2}}{2}\Big)\pi\Lambda-\frac{\pi}{12}\frac{\lambda^{2}}{\lambda_{m}}-\frac{\pi}{3}\lambda_{m}+2\pi\xi\lambda_{m} \nn\\
&+&\frac{2\pi}{3}\lambda_{m}^{3}  \bigg].
\end{eqnarray}
and
\begin{eqnarray}\label{delta22}
\mathcal{T}_{22}^{(2)} &=& \Omega^{2}(\tau)\frac{H^{3}}{4\pi^{2}}\bigg[
\frac{\pi}{3}\Lambda^{3}+\Big(-\frac{1}{8}+\xi-\frac{1}{8}\lambda^{2}-\frac{\lambda_{m}^{2}}{2}\Big)\pi\Lambda+\frac{\pi}{12}\frac{\lambda^{2}}{\lambda_{m}}-\frac{\pi}{3}\lambda_{m}+2\pi\xi\lambda_{m} \nn\\
&+&\frac{2\pi}{3}\lambda_{m}^{3}  \bigg],
\end{eqnarray}
For the only nonvanishing off-diagonal components, we obtain counterterms to second order adiabatic
\begin{equation}\label{delta01}
\mathcal{T}_{01}^{(2)}=\mathcal{T}_{10}^{(2)}=\Omega^{2}(\tau)\frac{H^{3}}{4\pi^{2}}\Big(\pi \lambda\Big)\Lambda^{2}.
\end{equation}
\subsection{Why subtraction stops at second adiabatic order} \label{Why:subtraction}
The required subtraction order can be determined from a dimensional analysis of the ultraviolet behavior of the energy-momentum tensor. In three spacetime dimensions, the energy-momentum tensor involves integrals over two-dimensional momentum space,
\begin{equation}
\langle T_{\mu\nu} \rangle \sim \int d^2 k \, \mathcal{T}_{\mu\nu}(k).
\end{equation}
At large momentum, the adiabatic expansion yields an asymptotic series in inverse powers of $k$. The leading contribution behaves as $\mathcal{T}_{\mu\nu}(k) \sim k$, producing a cubic divergence $\sim \Lambda^3$, which is removed by the zeroth-order adiabatic term.
The next-to-leading contribution behaves as $\mathcal{T}_{\mu\nu}(k) \sim k^{-1}$, leading to a linear divergence $\sim \Lambda$, which is canceled by the second-order adiabatic term.
Higher-order terms decay faster, $\mathcal{T}_{\mu\nu}(k) \sim k^{-3}$ or stronger, and therefore give finite contributions in three dimensions. This demonstrates that adiabatic subtraction up to second order is sufficient to remove all ultraviolet divergences.
This expectation is further supported by the numerical analysis presented below. For illustration, representative results for $T_{00}(\Lambda)$ are presented in Appendix~\ref{app:uv}. The numerical fit yields
\begin{equation}
T_{00}(\Lambda) = a \Lambda^3 + b \Lambda + \mathcal{O}(\Lambda^{-1}),
\end{equation}
in agreement with the analytical expectation. Moreover,
\begin{equation}
\frac{\Delta T_{00}}{\Lambda^3} \to 0 \quad \text{as} \quad \Lambda \to \infty,
\end{equation}
confirming the absence of logarithmic divergences and validating the second-order truncation of the adiabatic expansion.
\subsection{Renormalized tensor and Ward identity}
The adiabatic regularization procedure is carried out by subtracting the counterterms \eqref{delta00}-\eqref{delta01} from the corresponding unregularized in-vacuum expectation values \eqref{vinemt00}, \eqref{vinemt11}, \eqref{vinemt22}, and \eqref{offdiagg}. Thus we obtain our final expression for the timelike component of the regularized energy-momentum tensor
\begin{flalign}\label{reg00}
T_{00}&=\big\langle\mathrm{in}\big|T_{00}\big|\mathrm{in}\big\rangle-\mathcal{T}_{00}^{(2)}  \nn\\
&= \Omega^{2}(\tau)\frac{H^{3}}{4\pi^{2}}\bigg\{-
\frac{\pi}{12}\frac{\lambda^{2}}{\lambda_{m}}-\frac{\pi}{3}\lambda_{m}+2\pi\xi\lambda_{m}+\frac{2\pi}{3}\lambda_{m}^{3}
-\frac{3}{32}i \pi+\frac{5}{12}i \pi \gamma^{2}-\frac{1}{2}\pi^{2} \gamma^{2} \nn\\
&-\frac{1}{6} i\pi \gamma^{4}+\frac{1}{2} \pi^{2} \gamma^{4}-\frac{7}{24} i \pi \lambda^{2}-\frac{1}{2} i\pi \gamma^{2}\lambda^{2}-\frac{5}{16} i \pi\lambda^{4} + \Big(-\frac{13}{24}+\frac{1}{6}\gamma^{2}+4\xi+\frac{1}{4}\lambda^{2}\Big)\nn\\
&\times\pi \gamma\cot\big(2\pi\gamma\big)+\Big(\frac{2}{3}-\frac{2}{3}\gamma^{2}+4\xi-\lambda^{2}\Big) \pi\gamma \csc\big(2\pi\gamma\big){\mathrm{I}}_{0}\big(2\pi\lambda\big)+\frac{1}{2}\gamma\lambda \csc\big(2\pi\gamma\big) {\mathrm{I}}_{1}\big(2\pi\lambda\big) \nn\\
&+i\csc\big(2\pi\gamma\big)\int_{-1}^{+1}\frac{dr}{\sqrt{1-r^{2}}}B_{0r}\nn\\
&\bigg[\Big(e^{2\pi\lambda r}+e^{-2i\pi\gamma}\Big)
\psi\Big(\frac{1}{2}-\gamma+i\lambda r\Big)
-\Big(e^{2\pi\lambda r}+e^{2i\pi\gamma}\Big)\psi\Big(\frac{1}{2}+\gamma+i\lambda r\Big)\bigg]. &&
\end{flalign}
We obtain our final expressions for the diagonal spacelike components of the regularized energy-momentum tensor
\begin{flalign}\label{reg11}
T_{11}&=\big\langle\mathrm{in}\big|T_{11}\big|\mathrm{in}\big\rangle-\mathcal{T}_{11}^{(2)} \nn\\
&\Omega^{2}(\tau)\frac{H^{3}}{4\pi^{2}}\bigg\{
\frac{\pi}{12}\frac{\lambda^{2}}{\lambda_{m}}+\frac{\pi}{3}\lambda_{m}-2\pi\xi\lambda_{m} -\frac{2\pi}{3}\lambda_{m}^{3}
-\frac{3}{32} i\pi+\frac{5}{12}i \pi \gamma^{2}-\frac{1}{2}\pi^{2} \gamma^{2}\nn\\
&-\frac{1}{6} i\pi \gamma^{4} +\frac{1}{2} \pi^{2} \gamma^{4}+\frac{3}{8}i\pi \xi-\frac{5}{3}i\pi\xi \gamma^{2}+ 2\pi^{2}\xi\gamma^{2}+\frac{2}{3} i\pi\gamma^{4}\xi-2\pi^{2}\gamma^{4}\xi-\frac{7}{24} i\pi\lambda^{2}-\frac{1}{2} i\pi\lambda^{2}\gamma^{2}  \nn\\
&+\frac{7}{6} i\pi\xi\lambda^{2}+2 i\pi\xi\lambda^{2}\gamma^{2}-\frac{5}{16}i\pi\lambda^{4}+\frac{5}{4}i\pi\xi\lambda^{4}+\Big(-\frac{5}{24}-\frac{1}{6}\gamma^{2}+\frac{9}{2}\xi-2\gamma^{2}\xi-24\xi^{2}-\frac{1}{8}\lambda^{2}
\nn\\
&-3 \xi \lambda^{2}+\lambda_{m}^{2}-4 \xi \lambda_{m}^{2}\Big) \pi \gamma\cot\big(2\pi\gamma\big)+\Big(-\frac{15}{4}\frac{1}{\pi^{2}}+\frac{5}{3}-\frac{5}{3}\gamma^{2}-\frac{1}{3}\xi+\frac{4}{3}\xi\gamma^{2}-24\xi^{2}-2\lambda^{2} \nn\\
&+4\xi\lambda^{2}+\lambda_{m}^{2}-4\xi \lambda_{m}^{2}\Big) \pi\gamma\csc\big(2\pi\gamma\big){\mathrm{I}}_{0}\big(2\pi\lambda\big)+\Big(-\frac{2}{3\lambda}+\frac{15}{4}\frac{1}{\pi^{2}\lambda}+\frac{2}{3}\frac{\gamma^{2}}{\lambda}+3\lambda-4\xi \lambda\Big) \nn\\
&\times \gamma \csc\big(2\pi\gamma\big) {\mathrm{I}}_{1}\big(2\pi\lambda\big) +i\csc\big(2\pi\gamma\big)\int_{-1}^{+1}\frac{dr}{\sqrt{1-r^{2}}}B_{1r} \nn\\
&\bigg[\Big(e^{2\pi\lambda r}+e^{-2i\pi\gamma}\Big)
\psi\Big(\frac{1}{2}-\gamma+i\lambda r\Big)
-\Big(e^{2\pi\lambda r}+e^{2i\pi\gamma}\Big)\psi\Big(\frac{1}{2}+\gamma+i\lambda r\Big)\bigg] ,&&
\end{flalign}
and

\begin{flalign}\label{reg22}
T_{22}&=\big\langle\mathrm{in}\big|T_{22}\big|\mathrm{in}\big\rangle-\mathcal{T}_{22}^{(2)} \nn\\
&\Omega^{2}(\tau)\frac{H^{3}}{4\pi^{2}}\bigg\{
-\frac{\pi}{12}\frac{\lambda^{2}}{\lambda_{m}}+\frac{\pi}{3}\lambda_{m}-2\pi\xi\lambda_{m}-\frac{2\pi}{3}\lambda_{m}^{3}
-\frac{3}{32} i\pi+\frac{5}{12}i \pi \gamma^{2} \nn\\
&-\frac{1}{2}\pi^{2} \gamma^{2}-\frac{1}{6} i\pi \gamma^{4}+\frac{1}{2} \pi^{2} \gamma^{4}+\frac{3}{8}i\pi \xi-\frac{5}{3}i\pi\xi \gamma^{2}+ 2\pi^{2}\xi\gamma^{2}+\frac{2}{3} i\pi\gamma^{4}\xi-2\pi^{2}\gamma^{4}\xi-\frac{7}{24} i\pi\lambda^{2} \nn\\
&-\frac{1}{2} i\pi\lambda^{2}\gamma^{2} +\frac{7}{6} i\pi\xi\lambda^{2}+2 i\pi\xi\lambda^{2}\gamma^{2}-\frac{5}{16}i\pi\lambda^{4}+\frac{5}{4}i\pi\xi\lambda^{4}+\Big(-\frac{5}{24}-\frac{1}{6}\gamma^{2}+\frac{9}{2}\xi-2\gamma^{2}\xi
\nn\\
&-24\xi^{2}+\frac{1}{8}\lambda^{2}-3 \xi \lambda^{2}+\lambda_{m}^{2}-4 \xi \lambda_{m}^{2}\Big) \pi \gamma\cot\big(2\pi\gamma\big)+\Big(\frac{15}{4}\frac{1}{\pi^{2}}+\frac{1}{3}-\frac{1}{3}\gamma^{2}-\frac{1}{3}\xi+\frac{4}{3}\xi\gamma^{2}\nn\\
&-24\xi^{2}-\lambda^{2}+4\xi\lambda^{2}+\lambda_{m}^{2}-4\xi \lambda_{m}^{2}\Big) \pi\gamma \csc\big(2\pi\gamma\big){\mathrm{I}}_{0}\big(2\pi\lambda\big)+\Big(\frac{2}{3\lambda}-\frac{15}{4}\frac{1}{\pi^{2}\lambda}-\frac{2}{3}\frac{\gamma^{2}}{\lambda}-\frac{3}{2}\lambda \nn\\
&-4\xi \lambda\Big)\gamma \csc\big(2\pi\gamma\big){\mathrm{I}}_{1}\big(2\pi\lambda\big) +i\csc\big(2\pi\gamma\big)\int_{-1}^{+1}\frac{dr}{\sqrt{1-r^{2}}}B_{2r}\nn\\
&\bigg[\Big(e^{2\pi\lambda r}+e^{-2i\pi\gamma}\Big)
\psi\Big(\frac{1}{2}-\gamma+i\lambda r\Big)
-\Big(e^{2\pi\lambda r}+e^{2i\pi\gamma}\Big)\psi\Big(\frac{1}{2}+\gamma+i\lambda r\Big)\bigg]. &&
\end{flalign}
The nonvanishing regularized expectation values of off-diagonal components, given by 
\begin{equation}\label{reg01}
T_{01}=T_{10}=\big\langle\mathrm{in}\big|T_{10}\big|\mathrm{in}\big\rangle-\mathcal{T}_{10}^{(2)}=0.
\end{equation}
It is not surprising that the off-diagonal component $T_{01}$ is non-vanishing before renormalization, since the applied electric field explicitly breaks spatial symmetry. However, after renormalization the flux becomes exactly zero. This cancellation arises because oppositely directed modes, corresponding to positive and negative charges, contribute equally to the momentum density with opposite signs. As a result, the renormalized off-diagonal component vanishes, indicating the absence of a net energy flux despite the presence of particle production and vacuum polarization effects.

We work with the metric signature $(+,-,-)$, so that $T_{00}$ denotes the covariant time–time component of the energy–momentum tensor in comoving coordinates. 
The physical energy density measured by a comoving observer is 
\begin{equation}
T_{\mu\nu}u^\mu u^\nu = \frac{T_{00}}{\Omega^2},
\end{equation}

where $u^\mu=\Omega^{-1}(1,0,0)$ is the observer four-velocity in conformal coordinates and indices are lowered with the background metric.
Because the scalar field interacts with an external classical electromagnetic background, the renormalized energy-momentum tensor is not covariantly conserved in isolation, but instead satisfies the Ward identity

\begin{equation}
\nabla_\mu T^{\mu\nu}_{\rm ren} = F^{\nu\lambda} J^{\rm ren}_{\lambda}.
\end{equation}
This relation shows that the energy density is not conserved independently, but can exchange energy with the external electric field. This will play an important role in determining the sign of $T_{00}$.

\section{Internal checks on the renormalization}
\label{sec:checks}

\subsection{Ultraviolet structure and numerical verification}
In this section, we present internal consistency checks of the renormalized energy--momentum tensor, focusing on its ultraviolet (UV) structure and selected numerical evaluations. The purpose of this analysis is not to provide an independent derivation of the tensor, but rather to verify that the adiabatic regularization procedure correctly reproduces the expected divergence structure and yields finite, well-defined results.
At the level of the bare expectation values, the energy--momentum tensor exhibits the standard UV divergences characteristic of scalar quantum field theory in curved spacetime. In three dimensions, these divergences appear as cubic and linear contributions in the momentum cutoff.
We explicitly verify that the adiabatic subtraction terms remove these divergences and that no additional logarithmic UV contributions arise. This provides a nontrivial check that the subtraction scheme is implemented consistently and correctly captures the full UV structure of the theory.
To further support this conclusion, we numerically evaluate the difference between the bare and adiabatic subtraction terms defining the regulated mode integrals for the renormalized components of the energy--momentum tensor. In particular, we verify that the resulting expressions converge to stable, cutoff-independent values.
We emphasize that this numerical analysis serves only as an internal consistency check of the regularization procedure. It does not constitute an independent derivation of the renormalized energy--momentum tensor, nor does it provide a comparison with alternative renormalization schemes such as point-splitting.
Similarly, the Bogoliubov analysis presented later in this work captures only the particle-production contribution and should be regarded as a complementary diagnostic tool. It does not provide a quantitative reconstruction or cross-check of the full energy--momentum tensor, which also contains state-independent vacuum polarization contributions.
Additional details regarding the numerical implementation and convergence tests are provided in Appendix~\ref{app:uv}.

\section{Physical behavior of the renormalized tensor}
\label{sec:behavior}

\subsection{General numerical behavior}
\begin{figure}[H]\centering
\includegraphics[scale=0.7]{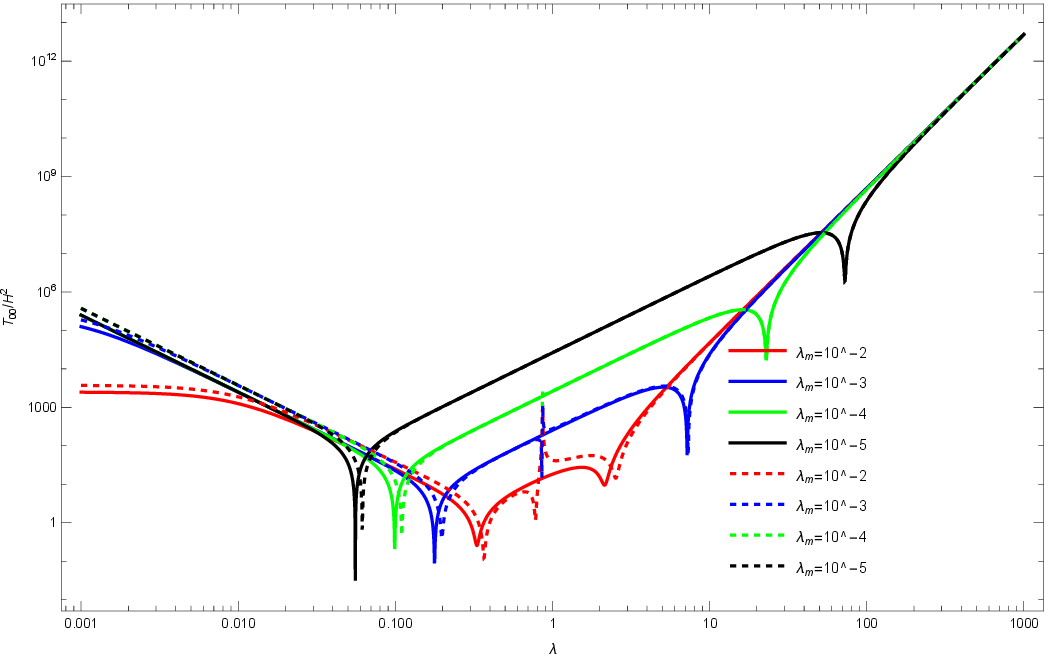}
\caption{Plot of the absolute value of  the $T_{00}$ component of the induced energy-momentum tensor is shown for the electric field parameter $\lambda=-eE/H^{2}$. The solid line $\xi=\frac{1}{8}$ and the dashed line $\xi=0$ correspond to different values of the mass parameter $\lambda_{m}=m/H$, as indicated. The scales on both axes are logarithmic.} \label{fig1}
\end{figure}

\begin{figure}[H]\centering
\includegraphics[scale=0.7]{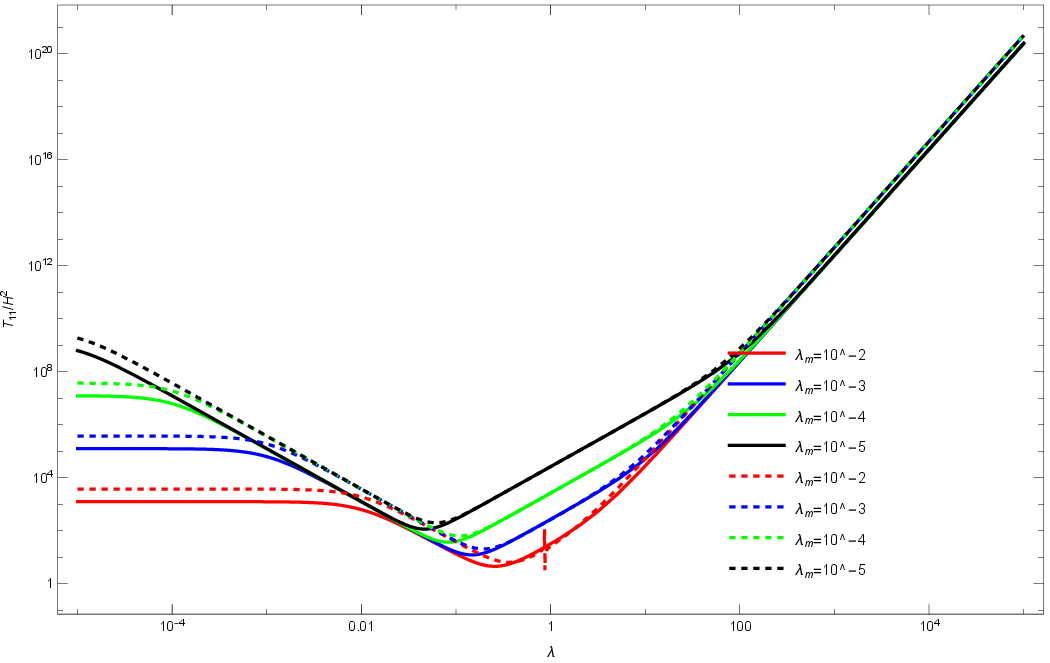}
\caption{Plot of the absolute value of the $T_{11}$ component of the induced energy-momentum tensor is shown for the electric field parameter $\lambda=-eE/H^{2}$. The solid line $\xi=\frac{1}{8}$ and the dashed line $\xi=0$ correspond to different values of the mass parameter $\lambda_{m}=m/H$, as indicated. The scales on both axes are logarithmic.} \label{fig2}
\end{figure}

\begin{figure}[H]\centering
\includegraphics[scale=0.7]{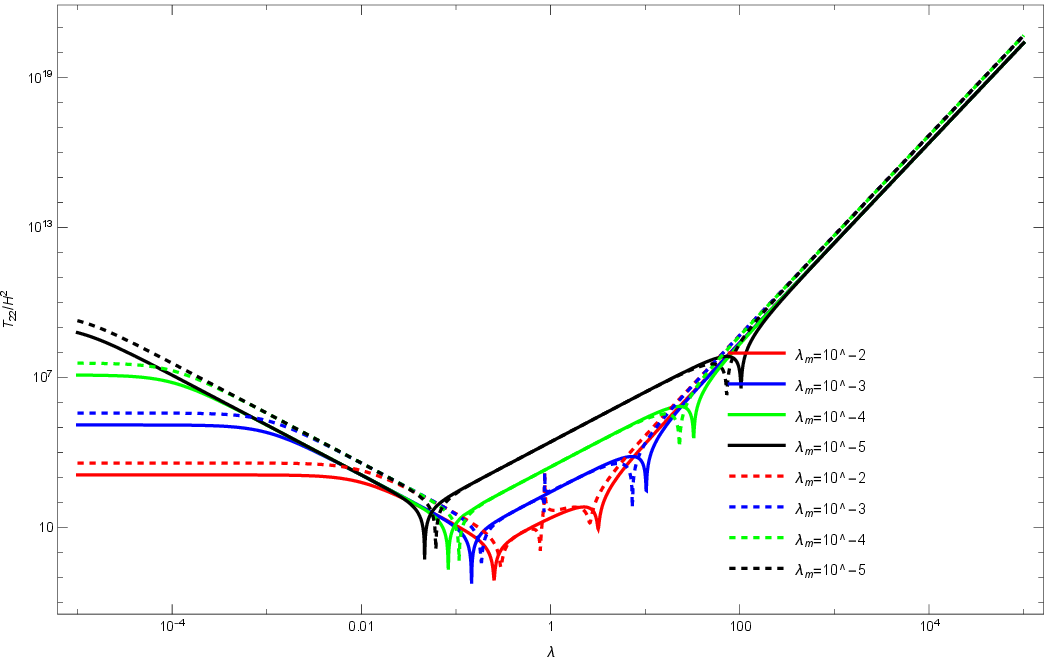}
\caption{Plot of the absolute value of the $T_{22}$ component of the induced energy-momentum tensor is shown for the electric field parameter $\lambda=-eE/H^{2}$. The solid line $\xi=\frac{1}{8}$ and the dashed line $\xi=0$ correspond to different values of the mass parameter $\lambda_{m}=m/H$, as indicated. The scales on both axes are logarithmic.} \label{fig3}
\end{figure}

In this subsection, we summarize the general numerical behavior of the renormalized energy--momentum tensor components as functions of the electric field strength, the mass parameter, and the curvature coupling, as illustrated in Figs.~\ref{fig1}-\ref{fig3}.
The figures display the absolute values of the components on logarithmic scales. Due to this choice, zeros of the functions—where the sign of the expressions changes—are shown as sharp cusp-like features. These should not be interpreted as physical singularities, but rather as artifacts of plotting absolute values on a logarithmic scale. The sign information is therefore not directly visible in these plots.
Overall, the numerical results exhibit a smooth and continuous dependence on the parameters, consistent with the analytical structure of the renormalized expressions. The magnitude of the energy--momentum tensor components generally increases with the electric field strength, reflecting the growth of the analytic vacuum polarization contribution, which dominates the leading behavior. 
Subleading, state-dependent contributions associated with particle production are not captured by the present analysis.
The dependence on the mass parameter indicates a suppression of the induced quantities for heavier fields, consistent with the decoupling of massive modes, while lighter fields show a stronger response, particularly in the strong-field regime, where the analytic vacuum polarization term becomes dominant.

The comparison between $\xi=0$ and $\xi=1/8$ illustrates the influence of curvature coupling. The conformal coupling $\xi=1/8$ leads to a weaker response to the electric field, which can be attributed to a partial cancellation of curvature-induced contributions in the energy--momentum tensor.

In the weak-field regime ($\lambda \ll 1$), the induced quantities are strongly suppressed and exhibit an approximate power-law dependence on $\lambda$, while in the strong-field limit ($\lambda \gg 1$) they show a rapid enhancement, showing a rapid enhancement with increasing field strength, driven by the analytic vacuum polarization contribution that dominates in this regime. 
This behavior is only qualitatively reminiscent of strong-field phenomena such as the Schwinger effect, 
and should not be interpreted as arising from particle production.

We emphasize that the present figures are intended to illustrate qualitative trends. A more complete characterization, including the explicit sign structure of the components, would require complementary plots without absolute values.

\subsection{Strong-field asymptotics}
In the strong electric field regime $\lambda \gg \max(1,\lambda_m,\xi)$, it is appropriate to examine the asymptotic behavior of the renormalized energy--momentum tensor in the limit $\lambda \to \infty$. By expanding Eqs.~(\ref{reg00})--(\ref{reg22}) at large $\lambda$ with $\lambda_m$ fixed, we identify the leading contributions to the tensor components.

\begin{equation}\label{strong}
T_{00}=-T_{11}=T_{22}\simeq\Omega^{2}\frac{H^{3}}{4\pi^{2}}\Big(-\frac{\pi}{12}\frac{\lambda^{2}}{\lambda_{m}}\Big).
\end{equation}

The leading term in Eq.~\eqref{strong} scales quadratically with the electric field strength and represents an analytic contribution to the renormalized energy--momentum tensor. This behavior originates from vacuum polarization and reflects the response of the quantum vacuum to the external field.
It is important to emphasize that Eq.~\eqref{strong} should not be interpreted as the energy density associated with particle production.
The scaling $\lambda^2/\lambda_m$ arises from the vacuum polarization part of the renormalized tensor, which is analytic in the electric field.
This distinguishes it from non-analytic contributions, which are typically associated with particle production in related contexts (see Sec.~\ref{sec:bogoliubov}).
The negative sign of $T_{00}$ in Eq.~\eqref{strong} does not signal any inconsistency. Rather, it reflects the structure of the renormalized energy--momentum tensor in the presence of an external field and should be interpreted within the framework of the corresponding Ward identity discussed earlier.
Subleading contributions, which are not captured by Eq.~\eqref{strong}, may contain additional structure. However, these terms are parametrically suppressed in the strong-field limit and do not affect the dominant asymptotic behavior identified here.
Finally, the dependence on $\lambda_m$ indicates an enhanced response for smaller masses, consistent with the trends observed in the numerical analysis.

\subsection{Infrared and light-mass asymptotics} \label{sec:ir}
The power-law infrared divergence as $m \to 0$ is a genuine dynamical effect and persists independently of trace considerations.
The infrared enhancement observed in the small-mass limit originates from the well-known amplification of long-wavelength modes in de Sitter spacetime. 
This behavior is independent of ultraviolet renormalization issues and should not be interpreted as being related to the absence or presence of a Weyl anomaly. 
It should also not be directly interpreted as arising from particle production, but rather reflects the dynamical sensitivity of light scalar fields to the expanding background.
A Taylor series expansion of the expressions \eqref{reg00}-\eqref{reg22} around $\lambda_{m}=0$, $\lambda=0$, and $\xi=0$ is appropriate for finding an approximate behavior of the induced energy-momentum tensor in the infrared regime with $\lambda_{m}\ll 1,\,\lambda\ll 1$ and $\xi=0$. The dominant terms in the expansions of \eqref{reg00}-\eqref{reg22}  are
\begin{eqnarray}
T_{00} \label{infrared00} &\simeq&\Omega^{2}\frac{H^{3}}{4\pi^{2}}\bigg(-\frac{17}{32}-\frac{\pi}{12}\frac{\lambda^{2}}{\lambda_{m}} 
+\mathcal{O}\Big(\lambda^{2},\lambda_{m}^{2}\Big)\bigg), \label{irlw0} \\
T_{11} \label{infrared11} &\simeq&\Omega^{2}\frac{H^{3}}{4\pi^{2}}\bigg(-\frac{43}{96}+\frac{\pi}{12}\frac{\lambda^{2}}{\lambda_{m}}
+\mathcal{O}\Big(\lambda^{2},\lambda_{m}^{2}\Big)\bigg), \label{irlw1} \\
T_{22} \label{infrared22} &\simeq&\Omega^{2}\frac{H^{3}}{4\pi^{2}}\bigg(-\frac{91}{96}-\frac{\pi}{12}\frac{\lambda^{2}}{\lambda_{m}}
+\mathcal{O}\Big(\lambda^{2},\lambda_{m}^{2}\Big)\bigg). \label{irlw2}
\end{eqnarray}
It is noticeably the case that the asymptotic expansions  \eqref{infrared00}-\eqref{infrared22} diverge as $m^{-1}$ in the exactly massless case. 
 Equations \eqref{infrared00}-\eqref{infrared22} show that the leading terms of this expansion are proportional to $H^{3}$, with correction terms involving powers of $\lambda$ and $\lambda_m$. 
 The negative signs in $T_{00}$ and the spatial components suggest a decrease in the local energy density, reflecting the response of the quantum field to the external background.
Moreover, the presence of terms proportional to 
$\lambda^{2}/\lambda_{m}$ 
shows that the energy--momentum tensor diverges as $\lambda_m^{-1}$. Since $\lambda_m = m/H$, this corresponds to a linear divergence $\propto m^{-1}$  in the massless limit.
This behavior reveals an infrared divergence associated with the 
massless limit, which is a well-known feature of quantum field 
theories in curved spacetime~\cite{PhysRevD.9.341,Fulling1974}.  
Physically, this indicates that the vacuum fluctuations become increasingly dominant in the nearly massless case, 
and the renormalized energy–momentum tensor loses its finite character in the exact massless limit~\cite{kobayashi2014schwinger}.
This infrared divergence is therefore linear in $1/m$ and reflects the enhancement of long-wavelength modes in the light-mass regime.
Ultraviolet properties, such as the trace structure and possible anomaly contributions, 
are discussed separately and do not affect the infrared behavior analyzed here.

\subsection{Comparison with \texorpdfstring{$\mathrm{dS}_2$ and $\mathrm{dS}_4$}{dS2 and dS4} results}
It is instructive to compare the present three-dimensional results with the corresponding analyses in two and four spacetime dimensions, focusing on the specific structural features emphasized in this work.
First, in the strong-field regime, the leading contribution to the renormalized energy--momentum tensor in $\mathrm{dS}_3$ is analytic in the electric field and arises from vacuum polarization. This behavior is qualitatively similar to the results in $\mathrm{dS}_2$ and $\mathrm{dS}_4$, where the leading terms also correspond to state-independent contributions. However, in contrast to the non-analytic particle-production contributions associated with Bogoliubov coefficients, the dominant term identified here remains purely analytic.

Second, the infrared and light-mass behavior exhibits a marked dimensional dependence.  In the present three-dimensional case, the energy--momentum tensor diverges as $\lambda_m^{-1} \sim H/m$ in the massless limit, reflecting a strong infrared 
enhancement driven by long-wavelength modes. 
This behavior differs qualitatively from the structure found in two and four dimensions, where the massless limit typically involves weaker, often logarithmic dependencies on the mass, depending on the observable and the renormalization scheme.  In particular, such contributions are not characterized by a simple power-law divergence 
of the type found here.

Finally, the role of the trace anomaly differs across dimensions. In even-dimensional spacetimes such as $\mathrm{dS}_2$ and $\mathrm{dS}_4$, the renormalized energy--momentum tensor generally exhibits a nonvanishing trace associated with the conformal anomaly. In contrast, in three spacetime dimensions no such anomaly is present, and the trace structure does not contribute to the infrared behavior discussed above. This distinction further supports the interpretation that the infrared divergence identified here is a genuine dynamical effect, independent of ultraviolet trace properties.

Overall, this comparison highlights that while the qualitative features of vacuum polarization persist across dimensions, the detailed scaling behavior and the role of trace contributions are strongly dimension-dependent.

\section{Renormalized trace in the conformal massless limit}
\label{sec:trace}

\subsection{Expression for the trace}
The trace of the induced energy–momentum tensor
$T$ is calculated by contracting the metric \eqref{metric:flrw} with the tensor components presented in Eqs.~\eqref{reg00}, \eqref{reg11}, and~\eqref{reg22}.
\begin{flalign}\label{trace}
T = g^{\mu\nu}T_{\mu\nu}&= \nn\\
&\frac{H^{3}}{4\pi^{2}}\bigg\{
\frac{3}{32}i\pi-\frac{5}{12}i \pi \gamma^{2}+\frac{1}{2}\pi^{2} \gamma^{2}+\frac{1}{6} i\pi \gamma^{4}-\frac{1}{2} \pi^{2} \gamma^{4}-\frac{3}{4}i\pi \xi+\frac{10}{3}i\pi\xi \gamma^{2}-4\pi^{2}\xi\gamma^{2}\nn\\
&-\frac{4}{3} i\pi\gamma^{4}\xi +4\pi^{2}\gamma^{4}\xi+\frac{7}{24} i\pi\lambda^{2} 
+\frac{1}{2} i\pi\lambda^{2}\gamma^{2} -\frac{7}{3} i\pi\xi\lambda^{2}-4 i\pi\xi\lambda^{2}\gamma^{2}+\frac{5}{16}i\pi\lambda^{4}\nn\\
&-\frac{5}{2}i\pi\xi\lambda^{4}-\frac{\pi}{12}\frac{\lambda^{2}}{\lambda_{m}}-\pi \lambda_{m}+6\pi\xi\lambda_{m}+2\pi\lambda_{m}^{3}+\Big(-\frac{1}{8}+\frac{1}{2}\gamma^{2}-5\xi+4\gamma^{2}\xi
\nn\\
&+48\xi^{2}+\frac{1}{4}\lambda^{2}+6 \xi \lambda^{2}-2\lambda_{m}^{2}+8\xi \lambda_{m}^{2}\Big) \pi \gamma\cot\big(2\pi\gamma\big)+\Big(-\frac{4}{3}+\frac{4}{3}\gamma^{2}+\frac{14}{3}\xi\nn\\
&-\frac{8}{3}\xi\gamma^{2}+48\xi^{2}+2\lambda^{2}-8\xi\lambda^{2}-2\lambda_{m}^{2}+8\xi \lambda_{m}^{2}\Big) \pi\gamma \csc\big(2\pi\gamma\big){\mathrm{I}}_{0}\big(2\pi\lambda\big) \nn\\
&+\Big(-\lambda+8\xi\lambda\Big)\gamma \csc\big(2\pi\gamma\big){\mathrm{I}}_{1}\big(2\pi\lambda\big) +i\csc\big(2\pi\gamma\big)\int_{-1}^{+1}\frac{dr}{\sqrt{1-r^{2}}}\nn\\
&\Big(\frac{1}{4}r\lambda-3 r\xi\lambda+4 r\gamma^{2}\xi\lambda+24 r\xi^{2}\lambda+4 r\xi \lambda^{3}-r \lambda \lambda^{2}_{m}+ 4 r \xi \lambda \lambda^{2}_{m}\Big) \nn\\
&\bigg[\Big(e^{2\pi\lambda r}+e^{-2i\pi\gamma}\Big)
\psi\Big(\frac{1}{2}-\gamma+i\lambda r\Big)
-\Big(e^{2\pi\lambda r}+e^{2i\pi\gamma}\Big)\psi\Big(\frac{1}{2}+\gamma+i\lambda r\Big)\bigg]. &&
\end{flalign}
We find that the renormalized trace takes the following form:
\begin{equation} \label{Trace3D}
\lim_{\lambda \to 0}  \lim_{\xi \to \frac{1}{8}} \lim_{\lambda_{m} \to 0} T= 0
\end{equation}
It should be emphasized that in three spacetime dimensions there is no genuine Weyl anomaly. Accordingly, the quantity obtained here should be interpreted as the renormalized trace rather than as a trace anomaly.
In the conformally coupled massless limit, the renormalized trace vanishes, in agreement with the general expectation that no Weyl anomaly arises in odd-dimensional spacetimes. Any finite residual contributions that may appear in intermediate steps, or depend on the choice of subtraction scheme or the ordering of limits, should be regarded as scheme-dependent artifacts of the renormalization procedure rather than physical effects.
From a physical perspective, this result confirms the absence of conformal symmetry breaking for a massless, conformally coupled scalar field in three-dimensional de Sitter space.

\section{Bogoliubov analysis and its relation to the renormalized tensor}
\label{sec:bogoliubov}

\subsection{Bogoliubov coefficients and particle number}
To further assess the physical content of our results, we perform in this section a Bogoliubov analysis of particle production and compute the corresponding particle production energy density.
Since both sets of modes form complete orthonormal bases, see Eqs.~ \eqref{uinb}, \eqref{vinb}, \eqref{uoutb} and \eqref{voutb}  the out-modes can be expanded in terms of the in-modes as
\begin{equation}\label{Bog_expansion}
U_{\text{out},\mathbf{k}}(x)
=
\int \frac{d^{2}k'}{(2\pi)^{2}}
\left[
\alpha_{\mathbf{k}\mathbf{k}'}\,U_{\text{in},\mathbf{k}'}(x)
+
\beta_{\mathbf{k}\mathbf{k}'}\,V_{\text{in},\mathbf{k}'}(x)
\right].
\end{equation}
By virtue of orthonormality relations \eqref{orthonormb}, the Bogoliubov coefficients are determined by
\begin{equation}\label{Bog_coeff}
\alpha_{\mathbf{k}\mathbf{k}'}
=
\big( U_{\text{out},\mathbf{k}},\, U_{\text{in},\mathbf{k}'} \big),
\qquad
\beta_{\mathbf{k}\mathbf{k}'}
=
- \big( U_{\text{out},\mathbf{k}},\, V_{\text{in},\mathbf{k}'} \big),
\end{equation}
where the Bogoliubov coefficients satisfy the relations
\begin{equation}\label{Bog_relations2}
\int \frac{d^2 k}{(2\pi)^2}
\left[
\alpha^{*}_{\mathbf{k},\mathbf{k}'}\,\alpha_{\mathbf{k},\mathbf{k}''}
-
\beta_{\mathbf{k},\mathbf{k}'}\,\beta^{*}_{\mathbf{k},\mathbf{k}''}
\right]
=
(2\pi)^2 \delta^{(2)}(\mathbf{k}'-\mathbf{k}''),
\end{equation}
\begin{equation} \label{Bog_relations}
\int \frac{d^2 k}{(2\pi)^2}
\left[
\alpha^{*}_{\mathbf{k},\mathbf{k}'}\,\beta_{\mathbf{k},\mathbf{k}''}
-
\beta_{\mathbf{k},\mathbf{k}'}\,\alpha^{*}_{\mathbf{k},\mathbf{k}''}
\right]
=0.
\end{equation}
As a result of Eqs.~\eqref{phi-in-expansionb}, \eqref{phi-out-expansionb}, and \eqref{Bog_expansion}, the late-time annihilation operator
$a_{\text{out}\,\mathbf{k}}$ can be expressed in terms of the early-time
operators $a_{\text{in}\,\mathbf{k}}$ and $b_{\text{in}\,\mathbf{k}}$ through a
Bogoliubov transformation,
\begin{equation}\label{Bog_transform_op}
a_{\text{out},\mathbf{k}}
=
\int \frac{d^2 k'}{(2\pi)^2}
\left[
\alpha^{*}_{\mathbf{k},\mathbf{k}'}\, a_{\text{in},\mathbf{k}'}
-
\beta^{*}_{\mathbf{k},\mathbf{k}'}\, b^{\dagger}_{\text{in},\mathbf{k}'}
\right].
\end{equation}
Using  relation \eqref{Bog_transform_op} together with the in-vacuum state $|0\rangle_{\text{in}}$,
one can evaluate the expectation value of the particle number operator,
\begin{equation}\label{N_expect}
{}_{\text{in}}\!\left\langle 0 \right|
a^{\dagger}_{\text{out}\,\mathbf{k}} a_{\text{out}\,\mathbf{k}}
\left| 0 \right\rangle_{\text{in}}
=
\int \frac{d^2 k'}{(2\pi)^2}\,
\left|\beta_{\mathbf{k},\mathbf{k}'}\right|^2 .
\end{equation}
Therefore, if $\beta_{\mathbf{k},\mathbf{k}'} \neq 0$, particle creation occurs.

In order to compute the density of created pairs, an explicit form of the
Bogoliubov coefficients is required. These coefficients are obtained by
substituting the orthonormal mode functions given in Eqs.~ \eqref{uinb}, \eqref{vinb}, \eqref{uoutb} and \eqref{voutb} into
Eq.~\eqref{Bog_coeff}. One finds
\begin{equation}\label{Bog_coef1}
\alpha_{\mathbf{k}\mathbf{k}'} = (2\pi)^2 \delta^{(2)}(\mathbf{k}-\mathbf{k}')\,\alpha_k,
\qquad
\beta_{\mathbf{k}\mathbf{k}'} = (2\pi)^2 \delta^{(2)}(\mathbf{k}+\mathbf{k}')\,\beta_k .
\end{equation}
The explicit expressions for $\alpha_k$ and $\beta_k$ are
\begin{equation} \label{alpha_k_final}
\alpha_k
=
(2|\gamma|)^{1/2}\,
\frac{\Gamma(-2\gamma)}{\Gamma\!\left(\tfrac{1}{2}-\gamma-\kappa\right)}
\,e^{\frac{i\pi}{2}(\kappa-\gamma)} ,
\end{equation}
\begin{equation} \label{beta_k_final}
\beta_k
=
-\,i\,(2|\gamma|)^{1/2}\,
\frac{\Gamma(-2\gamma)}{\Gamma\!\left(\tfrac{1}{2}-\gamma+\kappa\right)}
\,e^{\frac{i\pi}{2}(\kappa+\gamma)} .
\end{equation}
The normalization condition $|\alpha_k|^2-|\beta_k|^2=1$ is satisfied. The
expected number of created pairs with comoving momentum $k$ in the in-vacuum
is determined by Eq.~\eqref{N_expect}. After a straightforward calculation, one obtains
\begin{equation} \label{beta_relation_final}
\left|\beta_{\mathbf{k},\mathbf{k}'}\right|^2
=
\Big((2\pi)^2 \, \delta^{(2)}(\mathbf{k}+\mathbf{k}')\Big)^{2}
\, \left|\beta_{\mathbf{k}}\right|^2 .
\end{equation}
with
\begin{equation}\label{beta_square_final}
|\beta_k|^2
= \frac{e^{-2\pi|\gamma|}+e^{-2\pi i \kappa}}
{2\sinh(2\pi|\gamma|)}.
\end{equation}
For positive–frequency modes, Eq.~\eqref{feq} can equivalently be expressed as
\begin{equation} \label{posi}
\frac{d^2 f(\tau)}{d\tau^2} + \omega^2 f(\tau) = 0,
\end{equation}
where the momentum dependent frequency is
\begin{equation}
\omega_k(\tau) = k^2 - \frac{2 e E k r}{H^2 \tau} + \frac{1}{\tau^2}
\left(\frac{m^2}{H^2} + \frac{e^2 E^2}{H^4} + 6 \bar{\xi} 
\right).
\end{equation}
The adiabatic regime is defined by the requirement that the time–dependence of the effective 
frequency $\omega_k(\tau)$ remains sufficiently slow during the evolution. More precisely, 
the adiabatic conditions impose
\begin{equation}
\frac{\dot{\omega}_k^{\,2}}{\omega_k^{4}} \ll 1,
\qquad
\frac{\ddot{\omega}_k}{\omega_k^{3}} \ll 1,
\label{adiabatic_conditions}
\end{equation}
where overdots denote derivatives with respect to the conformal time $\tau$. These inequalities 
ensure that the notion of particles is well defined and that the WKB approximation remains valid.
In the asymptotic past limit, $\tau \to -\infty$, the time–dependent frequency reduces to
\begin{equation}
\omega_k(\tau) \longrightarrow k,
\end{equation}
so that the above adiabatic conditions are automatically satisfied. Consequently, the quantum state approaches the standard Minkowski vacuum in this limit, which provides a consistent definition of the in–vacuum.
Under these conditions, particle production can be unambiguously characterized by the Bogoliubov coefficients, and the corresponding energy density of the produced particles is given by
\begin{equation} \label{rhodensity}
\rho_{\mathrm{Bog}}(k)
= \frac{1}{\Omega(\tau)^{2}} \int \frac{d^2 k}{(2\pi)^2}
\, {\omega}_k \, |\beta_k|^2,
\end{equation}

\subsection{Particle-production contribution to the energy density}
The energy density of the produced particles, as measured in this vacuum, is then given by
\begin{equation} \label{rhoans}
\rho(k) = \frac{1}{4\pi \sinh(2\pi|\gamma|)} \Big(e^{-2\pi |\gamma|} + I_0(2\pi \lambda ) \Big)
\int_{0}^{\infty} k^2\, dk. 
\end{equation}

The expression above represents the energy density carried by the produced particles in the adiabatic out-vacuum. 

The Bogoliubov method captures only the state-dependent particle-production contribution, 
whereas the full renormalized energy-momentum tensor also contains state-independent 
vacuum-polarization terms. Therefore, Eq.~\eqref{rhoansfin} should not be interpreted as a 
quantitative consistency check of Eq.~\eqref{strong}. Rather, it provides a complementary 
and partial description of the energy density, and differences in scaling are expected.
In the strong electric field regime, the Bogoliubov spectrum is controlled by the scale set 
by the electric field, with dominant contributions arising from modes with $k^2 \sim eE$. 
This implies that the relevant momentum scale is $k \sim \sqrt{eE}$. 
Accordingly, Eq.~\eqref{rhoans} provides the following scaling estimate for the energy density:
\begin{equation} \label{rhoansfin}
\rho(k)\sim\frac{1}{12\pi \sinh\left(2\pi |\gamma|\right)}
\left(
e^{-2\pi |\gamma|} + I_{0}(2\pi \lambda)
\right) \,(eE)^{3/2}.
\end{equation}
Therefore, this expression should not be interpreted as an exact result, but rather as a scaling estimate capturing the dominant contribution to the energy density from particle production. This behavior is analogous to the Schwinger effect in flat space, with curvature introducing an additional physical scale.
We performed a Bogoliubov analysis of particle production and computed the associated energy density. 
The Bogoliubov method captures only the state-dependent (particle-production) contribution, 
whereas the full renormalized energy–momentum tensor also contains state-independent vacuum polarization terms. 
In the strong-field regime, the behavior shown in Eq.~\eqref{strong} is dominated by the state-independent contribution 
arising from the adiabatic subtraction terms, which are not included in the Bogoliubov framework. 
Therefore, Eq.~\eqref{rhoansfin} should not be interpreted as a quantitative consistency check of Eq.~\eqref{strong}. 
Rather, the Bogoliubov analysis provides a complementary and partial description of the energy density, 
while differences in scaling are expected.

\section{Conclusion}
\label{sec:conclusion}
In this work, we present a renormalized energy--momentum tensor for a charged scalar field in three-dimensional de Sitter spacetime, evaluated in the Poincar\'e patch and consistent with the corresponding Ward identity, in the presence of a constant electric field using the adiabatic regularization scheme.
Our analysis reveals two distinct regimes. In the strong-field regime, $\lambda \gg 1$, the leading contribution to the renormalized tensor is dominated by an analytic vacuum-polarization term arising from adiabatic subtraction. In the infrared and light-mass regime, the energy--momentum tensor exhibits a pronounced dependence on the mass, with an enhancement associated with long-wavelength modes.
We verified the ultraviolet structure of the renormalized tensor, confirming the expected divergence pattern and the absence of additional logarithmic terms. 
The numerical evaluation provides an internal consistency check of the regularization procedure.
A Bogoliubov analysis was also performed to extract the particle-production contribution. This provides a complementary and partial description of the energy density, but does not constitute a quantitative reconstruction of the full renormalized energy--momentum tensor.
The present analysis does not include an independent derivation based on alternative renormalization schemes, which could provide an additional check of the results.
This analysis is limited to a scalar field in a fixed de Sitter background and does not include backreaction effects. Extensions to fields with spin, different couplings, or dynamical backgrounds would be of interest for future work.

\begin{appendices}

\section{Details of the ultraviolet numerical check}
\label{app:uv}
This appendix provides numerical details supporting the ultraviolet analysis presented in Sec.~\ref{Why:subtraction}.
We numerically evaluate the regulated expressions for 
$T_{00}(\Lambda)$. 
This numerical analysis provides a non-trivial consistency check 
of the ultraviolet divergence structure of the renormalized 
energy-momentum tensor.
For representative values of the ultraviolet cutoff $\Lambda$, we obtain the following asymptotic expression for the energy density $T_{00}(\Lambda)$:

\begin{table}[h]
\centering
\begin{tabular}{ccccc}
\hline
$\Lambda$ &
$T_{00}^{\mathrm{num}}(\Lambda)$ &
$T_{00}^{\mathrm{asymp}}(\Lambda)$ &
$\Delta T_{00}$ &
$\Delta T_{00}/\Lambda$ \\
\hline
20 & 8510.23 & 8432.17 & 78.06 & 3.90 \\
30 & 28473.75 & 28356.40 & 117.35 & 3.91 \\
40 & 67286.83 & 67130.20 & 156.63 & 3.92 \\
50 & 131232.65 & 131037.00 & 195.65 & 3.91 \\
60 & 226594.42 & 226359.00 & 235.42 & 3.92 \\
70 & 359655.44 & 359381.00 & 274.44 & 3.92 \\
80 & 536698.77 & 536385.00 & 313.77 & 3.92 \\
\hline
\end{tabular}
\caption{Comparison between the numerical momentum integral and the asymptotic analytical expression for the energy density $T_{00}(\Lambda)$.}
\end{table}

Therefore, the ultraviolet structure of the energy density takes the asymptotic form
\begin{equation} \label{T00U}
    T_{00}(\Lambda)
=
a_3 \Lambda^3
+
a_1 \Lambda
+
a_0
+
\mathcal{O}\!\left(\frac{1}{\Lambda}\right),
\end{equation}
where the leading cubic divergence predicted by Eq.~\eqref{delta00} is fully reproduced by the numerical evaluation, 
while the residual difference is consistent with a subleading linear contribution.
Moreover,
\begin{equation} \label{subleading}
\frac{\Delta T_{00}}{\Lambda^3}
\to 0
\quad \text{as} \quad
\Lambda \to \infty,
\end{equation}
with no numerical evidence for logarithmic contributions in the ultraviolet expansion.

This numerical analysis provides a non-trivial verification 
of the ultraviolet divergence structure of the renormalized 
energy-momentum tensor.
The numerical results confirm the expected cubic and linear
divergences and show no evidence of additional logarithmic terms.

A completely analogous analysis can be performed for $T_{11}$ and $T_{22}$, leading to the same ultraviolet structure. 
For brevity, we display here only the numerical results for $T_{00}$ as a representative case.

\section{Momentum integrals involving Whittaker functions}
\label{app:whittaker}
The explicit values of the coefficients $\mathcal{I}_{1},\mathcal{I}_{2},\ldots,\mathcal{I}_{7}$, which are appeared in Eqs.~\eqref{timeli}, \eqref{spaceli1} and \eqref{spaceli2}  are presented in the appendix. These coefficients are defined as
\begin{flalign}\label{def:I1}
\hspace{1cm}\mathcal{I}_{1} = e^{\pi\lambda r}\int_{0}^{\Lambda}dp p^{2} \Big|W_{\kappa,\gamma}\big(-2ip\big)\Big|^{2},&&
\end{flalign}
\begin{flalign}\label{def:I2}
\hspace{1cm}\mathcal{I}_{2} = e^{\pi\lambda r}\int_{0}^{\Lambda}dp p \Big|W_{\kappa,\gamma}\big(-2ip\big)\Big|^{2},&&
\end{flalign}
\begin{flalign}\label{def:I3}
\hspace{1cm}\mathcal{I}_{3} = e^{\pi\lambda r}\int_{0}^{\Lambda} dp\Big|W_{\kappa,\gamma}\big(-2ip\big)\Big|^{2},&&
\end{flalign}
\begin{flalign}\label{def:I4}
\hspace{1cm}\mathcal{I}_{4} = i e^{\pi\lambda r}\int_{0}^{\Lambda} dp p\Big(W_{\kappa,\gamma}\big(-2ip\big)W_{1-\kappa,\gamma}\big(2ip\big)-W_{-\kappa,\gamma}\big(2ip\big)W_{1+\kappa,\gamma}\big(-2ip\big)\Big),&&
\end{flalign}
\begin{flalign}\label{def:I5}
\hspace{1cm}\mathcal{I}_{5} = i e^{\pi\lambda r}\int_{0}^{\Lambda} dp \Big(W_{\kappa,\gamma}\big(-2ip\big)W_{1-\kappa,\gamma}\big(2ip\big)-W_{-\kappa,\gamma}\big(2ip\big)W_{1+\kappa,\gamma}\big(-2ip\big)\Big),&&
\end{flalign}
\begin{flalign}\label{def:I6}
\hspace{1cm}\mathcal{I}_{6} = e^{\pi\lambda r}\int_{0}^{\Lambda} dp \Big(W_{\kappa,\gamma}\big(-2ip\big)W_{1-\kappa,\gamma}\big(2ip\big)+W_{-\kappa,\gamma}\big(2ip\big)W_{1+\kappa,\gamma}\big(-2ip\big)\Big),&&
\end{flalign}
\begin{flalign}\label{def:I7}
\hspace{1cm}\mathcal{I}_{7} = e^{\pi\lambda r}\int_{0}^{\Lambda}dp W_{1+\kappa,\gamma}\big(-2ip\big)W_{1-\kappa,\gamma}\big(2ip\big),&&
\end{flalign}

The integrals given in \eqref{def:I1}-\eqref{def:I7} are of the same type as the integrals involving Whittaker functions that appear in the calculation of the induced current of the scalar field in $\mathrm{dS}_{2}$
\cite{frob2014schwinger} and $\mathrm{dS}_{4}$ \cite{kobayashi2014schwinger}. To evaluate these integrals, we use the same method introduced in~\cite{kobayashi2014schwinger}. Since the procedure for solving all integrals in \eqref{def:I1}-\eqref{def:I7}  is similar, we present only the detailed solution of the integral \eqref{def:I7}  as an example, and for the remaining integrals we merely state the results. 

To solve the integral \eqref{def:I7}, we begin by substituting the Mellin--Barnes integral representation of the Whittaker functions $W_{\kappa,\gamma}(z)$, given in equation \eqref{Mellin}. In this way, we obtain:
\begin{align}\label{intst}
\mathcal{I}_7 = C_7 \int_{-i\infty}^{+i\infty} \frac{ds}{2\pi i} \frac{dt}{2\pi i}
&\ \Gamma\left(\tfrac{1}{2}+\gamma+s\right)
\Gamma\left(\tfrac{1}{2}-\gamma+s\right)
\Gamma\left(-1+i\lambda r - s\right) \nonumber \\
&\ \times \Gamma\left(\tfrac{1}{2}+\gamma+t\right)
\Gamma\left(\tfrac{1}{2}-\gamma+t\right)
\Gamma\left(-1 - i\lambda r - t\right) \nonumber \\
&\ \times e^{\frac{i\pi}{2}(s-t)} 
\int_{0}^{\Lambda} (2p)^{-s-t} \, dp,
\end{align}
where
\begin{align}\label{C7}
C_7 = \frac{e^{\pi \lambda r}}{
\Gamma\left(-\tfrac{1}{2}+\gamma - i\lambda r\right)
\Gamma\left(-\tfrac{1}{2}-\gamma - i\lambda r\right)
\Gamma\left(-\tfrac{1}{2}+\gamma + i\lambda r\right)
\Gamma\left(-\tfrac{1}{2}-\gamma + i\lambda r\right)
}
\end{align}
In equation \eqref{intst}, the integral over $p$ is easily evaluated as
\begin{equation}\label{intp}
\int_{0}^{\Lambda} (2p)^{-s-t} \, dp = \frac{1}{2(1 - s - t)}\, (2\Lambda)^{1 - s - t}.
\end{equation}

In order for the result of the integral to vanish in the limit $\Lambda \to \infty$, as required, the integration contours for $s$ and $t$ must be chosen such that they satisfy the following condition:
\begin{equation} \label{cono}
\Re(s + t) < 1.
\end{equation}

To satisfy the condition~\eqref{cono}, we assume that the integration contours for $s$ and $t$ are chosen such that they always fulfill the following conditions:
\begin{equation} \label{cont}
\Re(s) < \frac{1}{2}, \qquad \Re(t) < \frac{1}{2}.
\end{equation}

Substituting the result of the integral~\eqref{intp} into equation~\eqref{intst}, we obtain
\begin{align} \label{I7f}
\mathcal{I}_7 = C_7 \int_{-i\infty}^{+i\infty} \frac{ds}{2\pi i}
\Gamma\left(\tfrac{1}{2}+\gamma+s\right)
\Gamma\left(\tfrac{1}{2}-\gamma+s\right)
\Gamma\left(-1+i\lambda r - s\right)
\int_{-i\infty}^{+i\infty} \frac{dt}{2\pi i}
\, f_r,s(t),
\end{align}
where, for fixed $s$ and $r$, the function $f_{r,s}(t)$ is defined as
\begin{align}\label{frst}
f_r,s(t) =\;&
\Gamma\left(\tfrac{1}{2}+\gamma+t\right)
\Gamma\left(\tfrac{1}{2}-\gamma+t\right)
\Gamma\left(-1 - i\lambda r - t\right)\,
e^{\frac{i\pi}{2}(s-t)}
\frac{(2\Lambda)^{\,1-s-t}}{2(1-s-t)}
\end{align}

To evaluate the integral in \eqref{I7f}, we use the residue theorem. 
The poles of the integrand arise from the Gamma functions 
$\Gamma\left(\tfrac{1}{2}+\gamma+t\right)$, 
$\Gamma\left(\tfrac{1}{2}-\gamma+t\right)$ 
and $\Gamma\left(-1-i\lambda r - t\right)$. 
Thus, the poles of $f_{r,s}(t)$ are given by

\begin{equation} \label{conditionone}
 t_L = -n - \frac{1}{2} \pm \gamma, \qquad
t_R = n-1 - i\lambda r,
\end{equation}

where $n = 0,1,2,\ldots$.
Also, there is a simple pole arising from the denominator in \eqref{frst}, which occurs at
\begin{equation} \label{ind}
    t_{R} = 1 - s.
\end{equation}

The poles $t_L$ arise from the Gamma functions 
$\Gamma\left(\tfrac{1}{2}+\gamma+t\right)$ and 
$\Gamma\left(\tfrac{1}{2}-\gamma+t\right)$, 
which lie on the left-hand side of the integration contour. 
On the other hand, the poles $t_R$ originate from 
$\Gamma\left(-1-i\lambda r - t\right)$, 
which lie on the right-hand side of the contour. 
We choose to close the integration contour in the right half-plane. 
Furthermore, we impose the following additional condition, together with condition~\eqref{conditionone}:
\begin{equation}\label{conditiontwo}
\Re(t) > -2.
\end{equation}
As a result, only the following poles contribute:
\begin{equation}\label{pole}
t_R = -1-i\lambda r,\;  - i\lambda r,\; 1- i\lambda r,\; \; 2- i\lambda r,\; 1 - s.
\end{equation}
In the limit $\Lambda \to \infty$, the non-vanishing residues remain. 
Therefore, the result of the integral over $t$ in Eq.~\eqref{I7f} can be written as

\begin{equation}\label{It_sum}
\int_{-i\infty}^{+i\infty} \frac{dt}{2\pi i}\, f_r,s(t) = h_{7,\mathrm{dep}(0)}(r,s)
+ h_{7,\mathrm{dep}(1)}(r,s)
+ h_{7,\mathrm{dep}(2)}(r,s)
+ h_{7,\mathrm{dep}(3)}(r,s)
+ h_{7,\mathrm{ind}}(r,s),
\end{equation}
where the residue at the pole $t_R =-1 - i\lambda r$ is given by
\begin{equation}\label{h70}
h_{7,\mathrm{dep}(0)}(r,s)
= i\, \Gamma\left(-\tfrac{1}{2}+\gamma - i\lambda r\right)
\Gamma\left(-\tfrac{1}{2}-\gamma - i\lambda r\right)
\frac{(2\Lambda)^{\,2 - s + i\lambda r}}{2(2 - s + i\lambda r)}
\, e^{\frac{i\pi}{2}(s + i\lambda r)},
\end{equation}
the residue at the pole $t_R = - i\lambda r$ is given by
\begin{equation}\label{h71}
h_{7,\mathrm{dep}(1)}(r,s)
= - \Gamma\!\left(\tfrac{1}{2} + \gamma - i\lambda r \right)
  \Gamma\!\left(\tfrac{1}{2} - \gamma - i\lambda r \right)
  \frac{(2\Lambda)^{\,1 - s + i\lambda r}}{2(1 - s + i\lambda r)}
  \, e^{\frac{i\pi}{2}(s + i\lambda r)},
\end{equation}
the residue at the pole $t_R = 1- i\lambda r$ is given by
\begin{equation}\label{h72}
h_{7,\mathrm{dep}(2)}(r,s)
= -\frac{i}{2}\,
\Gamma\!\left(\tfrac{3}{2} + \gamma - i\lambda r \right)
\Gamma\!\left(\tfrac{3}{2} - \gamma - i\lambda r \right)
\frac{(2\Lambda)^{-s + i\lambda r}}{2(-s + i\lambda r)},
\, e^{\frac{i\pi}{2}(s + i\lambda r)}
\end{equation}
the residue at the pole $t_R = 2- i\lambda r$ is given by
\begin{equation}\label{h73}
h_{7,\mathrm{dep}(3)}(r,s)
= \frac{1}{6}\,
\Gamma\!\left(\tfrac{5}{2} + \gamma - i\lambda r \right)
\Gamma\!\left(\tfrac{5}{2} - \gamma - i\lambda r \right)
\frac{(2\Lambda)^{-1 - s + i\lambda r}}{2(-1 - s + i\lambda r)},
\, e^{\frac{i\pi}{2}(s + i\lambda r)}
\end{equation}
the residue at the pole $t_R = 1-s$ is given by
\begin{equation}\label{h7ind}
h_{7,\mathrm{ind}}(r,s)
= -\frac{i}{2}\,
\Gamma\!\left(\tfrac{3}{2} + \gamma - s \right)
\Gamma\!\left(\tfrac{3}{2} - \gamma - s \right)
\Gamma\!\left(-2 - i\lambda r + s \right)
\, e^{i\pi s}
\end{equation}
Taking into account Eq.~\eqref{It_sum}, we rewrite Eq.~\eqref{I7f} as follows:
\begin{equation}  \label{I7sum}
\mathcal{I}_7 = I_{7,\mathrm{dep}(0)} + I_{7,\mathrm{dep}(1)} + I_{7,\mathrm{dep}(2)} + I_{7,\mathrm{dep}(3)} + I_{7,\mathrm{ind}},
\end{equation}
where
\begin{equation}
I_{7,\mathrm{dep}(0)} = C_7 \int_{-i\infty}^{+i\infty} \frac{ds}{2\pi i}
\, \Gamma\!\left(\tfrac{1}{2}+\gamma+s\right)
\Gamma\!\left(\tfrac{1}{2}-\gamma+s\right)
\Gamma(-1+i\lambda r - s)\,
h_{7,\mathrm{dep}(0)}(r,s),
\label{eq:I7_dep0}
\end{equation}

\begin{equation}
I_{7,\mathrm{dep}(1)} = C_7 \int_{-i\infty}^{+i\infty} \frac{ds}{2\pi i}
\, \Gamma\!\left(\tfrac{1}{2}+\gamma+s\right)
\Gamma\!\left(\tfrac{1}{2}-\gamma+s\right)
\Gamma(-1+i\lambda r - s)\,
h_{7,\mathrm{dep}(1)}(r,s),
\label{eq:I7_dep1}
\end{equation}

\begin{equation}
I_{7,\mathrm{dep}(2)} = C_7 \int_{-i\infty}^{+i\infty} \frac{ds}{2\pi i}
\, \Gamma\!\left(\tfrac{1}{2}+\gamma+s\right)
\Gamma\!\left(\tfrac{1}{2}-\gamma+s\right)
\Gamma(-1+i\lambda r - s)\,
h_{7,\mathrm{dep}(2)}(r,s),
\label{eq:I7_dep2}
\end{equation}

\begin{equation}
I_{7,\mathrm{dep}(3)} = C_7 \int_{-i\infty}^{+i\infty} \frac{ds}{2\pi i}
\, \Gamma\!\left(\tfrac{1}{2}+\gamma+s\right)
\Gamma\!\left(\tfrac{1}{2}-\gamma+s\right)
\Gamma(-1+i\lambda r - s)\,
h_{7,\mathrm{dep}(3)}(r,s),
\label{eq:I7_dep3}
\end{equation}

\begin{equation}
I_{7,\mathrm{ind}} = C_7 \int_{-i\infty}^{+i\infty} \frac{ds}{2\pi i}
\, \Gamma\!\left(\tfrac{1}{2}+\gamma+s\right)
\Gamma\!\left(\tfrac{1}{2}-\gamma+s\right)
\Gamma(-1+i\lambda r - s)\,
h_{7,\mathrm{ind}}(r,s).
\label{eq:I7_ind}
\end{equation}

The poles of the integrand of $I_{7,\mathrm{dep}(0)}$ \eqref{eq:I7_dep0} are given as follows:
\begin{equation}\label{poleI7}
s_L = -n - \frac{1}{2} \pm \gamma,
\qquad
s_R = n - 1 + i\lambda r,
\qquad n = 0,1,2,\ldots
\end{equation}
The left poles $s_L$ are given by the poles of 
$\Gamma\!\left(\frac{1}{2}+\gamma+s\right)$ and $\Gamma\!\left(\frac{1}{2}-\gamma+s\right)$, 
which lie on the left side of the integration contour. 
 The right poles $s_R$ arise from the poles of 
$\Gamma\!\left(-1+i\lambda - s\right)$, 
which lie on the right side of the integration contour and of the fraction on the right-hand side of Eq.~\eqref{h70}. 
We choose the contour such that it closes on the right half-plane.
Then, there are poles at
\begin{equation}
s_R = -1+i\lambda r,\; i\lambda r,\; 1+i\lambda r,\; 2+i\lambda r.
\end{equation}
In the limit $\Lambda \to \infty$, the residues at these poles do not vanish. From the above, the result of the integral Eq.~\eqref{eq:I7_dep0} is obtained as follows:
\begin{align}
I_{\text{7,dep(0)}}
&= \frac{4}{3}\Lambda^3
- i\left(-\frac{1}{2}+\gamma+i\lambda r\right)\left(-\frac{1}{2}-\gamma+i\lambda r\right)\Lambda^2 \notag \\[6pt]
&\quad - \frac{1}{2}
\left(\frac{1}{2}+\gamma+i\lambda r\right)
\left(-\frac{1}{2}+\gamma+i\lambda r\right)
\left(\frac{1}{2}-\gamma+i\lambda r\right)
\left(-\frac{1}{2}-\gamma+i\lambda r\right)\Lambda \notag \\[6pt]
&\quad - \frac{i}{12}
\left(\frac{3}{2}+\gamma+i\lambda r\right)
\left(\frac{1}{2}+\gamma+i\lambda r\right)
\left(-\frac{1}{2}+\gamma+i\lambda r\right) \notag \\
&\qquad \times
\left(\frac{3}{2}-\gamma+i\lambda r\right)
\left(\frac{1}{2}-\gamma+i\lambda r\right)
\left(-\frac{1}{2}-\gamma+i\lambda r\right) \notag \\[6pt]
&\quad \times \Bigg\{
\frac{1}{\frac{3}{2}+\gamma+i\lambda r}
+ \frac{1}{\frac{1}{2}+\gamma+i\lambda r}
+ \psi\!\left(\frac{1}{2}+\gamma+i\lambda r\right) \notag \\
&\qquad + \frac{1}{\frac{3}{2}-\gamma+i\lambda r}
+ \frac{1}{\frac{1}{2}-\gamma+i\lambda r}
+ \psi\!\left(\frac{1}{2}-\gamma+i\lambda r\right)
- \frac{11}{6} + \gamma_E
 \notag \\[6pt]
&\quad + \frac{i\pi}{2} - \ln(2\Lambda)\Bigg\},
\label{eq:I7dep0}
\end{align}
where $\gamma_E$ is the Euler--Mascheroni constant.

To obtain the result of the integrals $I_{\text{7,dep(1)}}$ to $I_{\text{7,dep(3)}}$, which are related to ~Eq.~\eqref{eq:I7_dep1}--Eq.~\eqref{eq:I7_dep3}, we proceed in a similar manner as for the integral $I_{\text{7,dep(0)}}$. Therefore, we have:
\begin{align}
I_{\text{7,dep(1)}}
&=i \left(-\tfrac{1}{2}+\gamma-i\lambda r\right)
\left(-\tfrac{1}{2}-\gamma-i\lambda r\right)\Lambda^2 \notag \\[6pt]
&\quad + 
\left(-\tfrac{1}{2}+\gamma+i\lambda r\right)
\left(-\tfrac{1}{2}-\gamma+i\lambda r\right)
\left(-\tfrac{1}{2}+\gamma-i\lambda r\right)
\left(-\tfrac{1}{2}-\gamma-i\lambda r\right)\Lambda \notag \\[6pt]
&\quad + \tfrac{i}{4}
\left(-\tfrac{1}{2}+\gamma-i\lambda r\right)
\left(-\tfrac{1}{2}-\gamma-i\lambda r\right)
\left(\tfrac{1}{2}+\gamma+i\lambda r\right)
\left(-\tfrac{1}{2}+\gamma+i\lambda r\right) \notag \\
&\qquad \times
\left(\tfrac{1}{2}-\gamma+i\lambda r\right)
\left(-\tfrac{1}{2}-\gamma+i\lambda r\right) \notag \\[6pt]
&\quad \times \Bigg\{
\frac{1}{\tfrac{1}{2}+\gamma+i\lambda r}
+ \,\psi\!\left(\tfrac{1}{2}+\gamma+i\lambda r\right) 
 \frac{1}{\tfrac{1}{2}-\gamma+i\lambda r}
+ \,\psi\!\left(\tfrac{1}{2}-\gamma+i\lambda r\right) \notag \\
&\qquad - \tfrac{3}{2} + \gamma_E
+ \tfrac{i\pi}{2} - \ln(2\Lambda)
\Bigg\},
\label{eq:I7dep1}
\end{align}

\begin{align}
I_{\text{7,dep(2)}}
&= \left(-\tfrac{1}{2}\right)
\left(\tfrac{1}{2}+\gamma-i\lambda r\right)
\left(-\tfrac{1}{2}+\gamma-i\lambda r\right)
\left(\tfrac{1}{2}-\gamma-i\lambda r\right)
\left(-\tfrac{1}{2}-\gamma-i\lambda r\right)\Lambda \notag \\[6pt]
&\quad - \tfrac{i}{4}
\left(\tfrac{1}{2}+\gamma-i\lambda r\right)
\left(-\tfrac{1}{2}+\gamma-i\lambda r\right)
\left(\tfrac{1}{2}-\gamma-i\lambda r\right)
\left(-\tfrac{1}{2}-\gamma-i\lambda r\right) \notag \\
&\qquad \times
\left(-\tfrac{1}{2}+\gamma+i\lambda r\right)
\left(-\tfrac{1}{2}-\gamma+i\lambda r\right) \notag \\[6pt]
&\quad \times \Bigg\{
\psi\!\left(\tfrac{1}{2}+\gamma+i\lambda r\right)
+ \psi\!\left(\tfrac{1}{2}-\gamma+i\lambda r\right)
- 1 + \gamma_E
+ \tfrac{i\pi}{2} - \ln(2\Lambda)
\Bigg\}
\label{eq:I7dep2},
\end{align}

\begin{align}
I_{\text{7,dep (3)}}
&= \frac{i}{12}
\left(\tfrac{3}{2}+\gamma-i\lambda r\right)
\left(\tfrac{1}{2}+\gamma-i\lambda r\right)
\left(-\tfrac{1}{2}+\gamma-i\lambda r\right) \notag \\
&\quad \times
\left(\tfrac{3}{2}-\gamma-i\lambda r\right)
\left(\tfrac{1}{2}-\gamma-i\lambda r\right)
\left(-\tfrac{1}{2}-\gamma-i\lambda r\right) \notag \\[6pt]
&\quad \times \Bigg\{
\psi\!\left(\tfrac{1}{2}+\gamma+i\lambda r\right)
- \frac{1}{-\tfrac{1}{2}+\gamma+i\lambda r} \notag \\
&\qquad + \psi\!\left(\tfrac{1}{2}-\gamma+i\lambda r\right)
- \frac{1}{-\tfrac{1}{2}-\gamma+i\lambda r}
+ \gamma_E + \frac{i\pi}{2} - \ln(2\Lambda)
\Bigg\}.
\label{eq:I7dep3}
\end{align}
 Using Eq.~\eqref{eq:I7_ind} and the properties of the Gamma function 
\begin{align}
\Gamma\!\left(\tfrac{1}{2}+z\right)\Gamma\!\left(\tfrac{1}{2}-z\right)
&= \frac{\pi}{\cos(\pi z)}, \\[6pt]
\Gamma(z)\Gamma(1-z)
&= \frac{\pi}{\sin(\pi z)}, \qquad z \neq 0,\pm1,\pm2,\ldots,
\end{align}
 the integral $I_{\text{7,ind}}$ can be rewritten as follows:
\begin{align}
I_{\text{7,ind}} = \left(-\frac{i}{2}\right)\pi^{3} C_{7}
\int_{-i\infty}^{+i\infty} \frac{ds}{2\pi i}\,
\frac{e^{i\pi s}}{\cos\pi(s+\gamma)\,\cos\pi(s-\gamma)} \notag \\
\times \frac{\left(s-\gamma-\tfrac{1}{2}\right)\left(s+\gamma-\tfrac{1}{2}\right)}
{(s-i\lambda r-2)(s-i\lambda r-1)(s-i\lambda r)(s-i\lambda r+1)} \notag \\
\times \frac{1}{\sin\pi(s-i\lambda r)}
\label{eq:Iind}
\end{align}
The poles of the integrand~\eqref{eq:Iind} are given as follows:
\begin{align}
s_R &=  n + i\lambda r, \qquad
s_L = - n + i\lambda r -2, \qquad n = 0,1,2,\ldots \label{eq:poles1} \\
s_R &= \frac{1}{2} + n + \gamma, \qquad
s_L = -\frac{1}{2} - n + \gamma, \qquad n = 0,1,2,\ldots \label{eq:poles2} \\
s_R &= \frac{1}{2} + n - \gamma, \qquad
s_L = -\frac{1}{2} - n - \gamma, \qquad n = 0,1,2,\ldots \label{eq:poles3}
\end{align}

As can be seen, the poles of $\sin\pi(s - i\lambda r)$ in Eq.~\eqref{eq:poles1} and the poles of $\cos\pi(s - \gamma)$ in Eq.~\eqref{eq:poles2}, as well as the poles of $\cos\pi(s + \gamma)$ in Eq.~\eqref{eq:poles3}, are given.
We consider the following expression:
\begin{equation}
U_{r}(s)= \frac{\left(s-\gamma-\tfrac{1}{2}\right)\left(s+\gamma-\tfrac{1}{2}\right)}
{(s-i\lambda r-2)(s-i\lambda r-1)(s-i\lambda r)(s-i\lambda r+1)},
\label{eq:B3}
\end{equation}
which can be rewritten as follows:
\begin{equation}
U_{r}(s) = g_r(s) - g_r(s-1),
\label{eq:B3_gr}
\end{equation}
where $ g_r(s)$ is given by
\begin{equation}
g_r(s) = \frac{-b(r) - c(r) - d(r)}{s - i\lambda r + 1}
-\frac{c(r) + d(r)}{s - i\lambda r}
- \frac{d(r)}{s - i\lambda r - 1},
\label{eq:gr}
\end{equation}
with
\begin{align}
a(r) &= \frac{\frac{9}{4} - \gamma^2 - 3 i r \lambda - r^2 \lambda^2}{6}, \\[6pt]
b(r) &= \frac{\frac{1}{4} - \gamma^2 - i r \lambda - r^2 \lambda^2}{2}, \\[6pt]
c(r) &= \frac{\frac{1}{4} - \gamma^2 + i r \lambda - r^2 \lambda^2}{2}, \\[6pt]
d(r) &= \frac{\frac{9}{4} - \gamma^2 + 3 i r \lambda - r^2 \lambda^2}{6}.
\end{align}
By substituting Eq.~\eqref{eq:B3_gr} into Eq.~\eqref{eq:Iind}, we obtain
\begin{equation}
I_{\text{7,ind}} = \left(-\frac{i}{2}\right)\pi^3 C_7
\int_{-i\infty}^{+i\infty} \frac{ds}{2\pi i}
\frac{e^{i\pi s}}{\cos\pi(s+\gamma)\,\cos\pi(s-\gamma)\,\sin\pi(s-i\lambda r)}
\left[ g_r(s) - g_r(s-1) \right].
\label{eq:Lind_final}
\end{equation}
To evaluate the integral~\eqref{eq:Lind_final}, we close the contour of integration in the left half-plane. Thus, it can be concluded that the contribution of the poles of $g_r(s)$ at $s=n$ is canceled by the contribution of the poles of $g_r(s-1)$ at $s=n+1$. Therefore, only the contributions of the following poles remain in the term $g_r(s)$.
\begin{equation} \label{poleleft}
    s_L = -\frac{1}{2}-\gamma,\quad -\frac{1}{2}+\gamma,\quad i\lambda_r-2,
\end{equation}
hence, the integral~~\eqref{eq:Lind_final} simplifies as follows.
\begin{equation}
I_{\text{7,ind}} = \left(-\frac{i}{2}\right)\pi^3 C_7
\int_{-i\infty}^{+i\infty} \frac{ds}{2\pi i}
\frac{e^{i\pi s}g_r(s) }{\cos\pi(s+\gamma)\,\cos\pi(s-\gamma)\,\sin\pi(s-i\lambda r)}
\label{eq:Lind_final2}
\end{equation}
In this way, the poles of the integrand in Eq.~\eqref{poleleft} are determined. Using the identity
\begin{equation}
\Gamma\!\left(\tfrac{1}{2}+iy\right)\Gamma\!\left(\tfrac{1}{2}-iy\right)
= \frac{\pi}{\cosh(\pi y)},
\end{equation}
$C_{7}$ in Eq.~\eqref{C7} can be rewritten as follows:
\begin{align}
C_7 &= \frac{1}{\pi^2} e^{\pi \lambda r}
\left( -\frac{1}{2} + \gamma - i\lambda r \right)
\left( -\frac{1}{2} - \gamma - i\lambda r \right) \nn\\
&\quad \times
\left( -\frac{1}{2} + \gamma + i\lambda r \right)
\left( -\frac{1}{2} - \gamma + i\lambda r \right)\nn \\
&\quad \times
\cos\pi(\gamma - i\lambda r)
\cos\pi(\gamma + i\lambda r)
\end{align}
Using the residue theorem for the poles of Eq.~\eqref{poleleft}, the result of the integral in Eq.~\eqref{eq:Lind_final2} is obtained as follows:
\begin{equation}\label{eq:Lind_finalfinal}
\begin{split}
I_{\text{7,ind}} =\;& \frac{1}{6}\gamma^3\lambda^3 
+ \frac{1}{24}(1+12\gamma^2) r\lambda 
+ \frac{1}{12}(1-4\gamma^2)\gamma \cot(2\pi\gamma) \\
&+ \frac{1}{12}(1-4\gamma^2)\gamma e^{2\pi\lambda r}\csc(2\pi\gamma) \\
&+ \left(-\frac{5}{256}i + \frac{91}{576}i\gamma^2 - \frac{47}{144}i\gamma^4 + \frac{1}{36}i\gamma^6 \right) \\
&+ \left(-\frac{1}{96} + \frac{1}{12}\gamma^2 - \frac{1}{6}\gamma^4 \right)\gamma\lambda \\
&+ \left(-\frac{89}{576} - \frac{5}{8} i\gamma^2 + \frac{1}{12}i\gamma^4 \right) r^{2}\lambda^2 \\
&+ \left(-\frac{5}{12} \right)\gamma^3\lambda^3 \\
&+ \left(-\frac{43}{144}i + \frac{1}{12}i\gamma^2 \right)\gamma^4\lambda^4 
- \frac{1}{6}\gamma^5\lambda^5 + \frac{1}{36}i\gamma^6\lambda^6.
\end{split}
\end{equation}
Finally,by substituting Eqs.~\eqref{eq:I7dep0}, \eqref{eq:I7dep1}, \eqref{eq:I7dep2}, \eqref{eq:I7dep3} and \eqref{eq:Lind_finalfinal} into Eq.~\eqref{I7sum}, we obtain
\begin{flalign}\label{endi7}
\hspace{0.5cm}	
\mathcal{I}_{7} &=\frac{4\Lambda^{3}}{3}-2\lambda r\Lambda^{2}
-\frac{3i}{32}+\frac{5i}{12}\gamma^{2}-\frac{i}{6}\gamma^{4}
-\frac{1}{3}\gamma^{4}\lambda r +\frac{2}{3}\gamma^{2}\lambda r+\frac{1}{48}\lambda r-\frac{7i}{12}\lambda^{2} r^{2}-i \gamma^{2} \lambda^{2} r^{2}\nn\\
&-\frac{2}{3}\gamma^{2} \lambda^{3} r^{3}-\frac{5 i}{6} \lambda^{4} r^{4}-\frac{1}{3} \lambda^{5} r^{5}+\frac{1}{12}\Big(1-4\gamma^{2}\Big)\gamma\Big(\cot\big(2\pi\gamma\big)+e^{2\pi\lambda r }\csc\big(2\pi\gamma\big)\Big).&&
\end{flalign}

To obtain the results of the integrals $\mathcal{I}_{1}$ through $\mathcal{I}_{6}$ from Eqs.~\eqref{def:I1}--\eqref{def:I6}, we follow a procedure similar to that used for the integral $\mathcal{I}_{7}$. The final results are thus obtained as follows:

\begin{flalign}\label{endi1}
\hspace{0.5cm}	
\mathcal{I}_{1} &=\frac{\Lambda^{3}}{3}+\frac{\lambda r}{2}\Lambda^{2}
+\frac{1}{8}\Big(4\gamma^{2}+12\lambda^{2}r^{2}-1\Big)\Lambda
+\frac{\lambda r}{8}\Big(12\gamma^{2}+20\lambda^{2}r^{2}-7\Big)\log\big(2\Lambda\big)
 \nn\\
&+\frac{95}{48}\lambda r-\frac{5\gamma^{2}}{4}\lambda r-\frac{3}{4}i \pi \gamma^{2} \lambda r+\frac{7}{16}i \pi  \lambda r-\frac{37}{12}\lambda^{3}r^{3}-\frac{5}{4} i\pi\lambda^{3} r^{3} \nn\\
&-\frac{\gamma}{6}\Big(4\gamma{^2}+15\lambda^{2}r^{2}-4\Big)
\Big(\cot\big(2\pi\gamma\big)+e^{2\pi\lambda r}\csc\big(2\pi\gamma\big)\Big)-\frac{i\lambda r}{16}\Big(12\gamma^{2}+20\lambda^{2}r^{2}-7\Big) \nn\\
&\times \csc\big(2\pi\gamma\big)\bigg[\Big(e^{2\pi\lambda r}+e^{-2i\pi\gamma}\Big)
\psi\Big(\frac{1}{2}-\gamma+i\lambda r\Big)
-\Big(e^{2\pi\lambda r}+e^{2i\pi\gamma}\Big)\psi\Big(\frac{1}{2}+\gamma+i\lambda r\Big)\bigg],&&
\end{flalign}

\begin{flalign}\label{endi2}
\hspace{0.5cm}
\mathcal{I}_{2} &=\frac{\Lambda^{2}}{2}+\lambda r\Lambda+\frac{1}{8}\Big(4\gamma^{2}+12\lambda^{2}r^2-1\Big)\log\big(2\Lambda\big)
-\frac{\gamma^{2}}{4}-\frac{7\lambda^{2}r^{2}}{4}+\frac{5}{16}-\frac{3}{4}i \pi \lambda^{2}r^{2}
 \nn\\
&-\frac{1}{4}i\pi\gamma^{2}+\frac{i}{16}\pi-\frac{3\gamma\lambda r}{2}\Big(\cot\big(2\pi\gamma\big)+e^{2\pi\lambda r }\csc\big(2\pi\gamma\big)\Big)-\frac{i}{16}\Big(4\gamma^{2}+12\lambda^{2}r^{2}
-1\Big) \nn\\
&\times\csc\big(2\pi\gamma\big)\bigg[\Big(e^{2\pi\lambda r}+e^{-2i\pi\gamma}\Big)\psi\Big(\frac{1}{2}-\gamma +i\lambda r\Big)-\Big(e^{2\pi\lambda r}+e^ {2i\pi\gamma}\Big)\psi\Big(\frac{1}{2}+\gamma+i\lambda r\Big)\bigg],&&
\end{flalign}

\begin{flalign}\label{endi3}
\hspace{1cm}\mathcal{I}_{3} &=
\Lambda+r\lambda\log\big(2\Lambda\big)-r\lambda-\frac{i}{2}r\lambda-\gamma\cot\big(2\pi\gamma\big)-\gamma\csc\big(2\pi\gamma\big)e^{2\pi\lambda r}
-\frac{i}{2}r\lambda\csc\big(2\pi\gamma\big) \nn\\
&\times\bigg[\Big(e^{2\pi\lambda r}+e^{-2\pi i\gamma}\Big)\psi\Big(\frac{1}{2}-\gamma+i\lambda r\Big)
-\Big(e^{2\pi\lambda r}+e^{2\pi i\gamma}\Big)\psi\Big(\frac{1}{2}+\gamma+i\lambda r\Big) \bigg],&&
\end{flalign}

\begin{flalign}\label{endi4}
\hspace{0.5cm}	
\mathcal{I}_{4} &=-\frac{4\Lambda^{3}}{3}-\lambda r\Lambda^{2}
-\frac{1}{4}\Big(4\gamma^{2}+4\lambda^{2}r^{2}-1\Big)\Lambda
+\lambda r\Big(-2\gamma^{2}-2\lambda^{2}r^{2}+\frac{3}{2}\Big)\log\big(2\Lambda\big)
 \nn\\
&-\frac{83}{24}\lambda r+\frac{3}{2}\gamma^{2}\lambda r-\frac{3}{4} i\pi \lambda r+ i\pi\gamma^{4} \lambda r+\frac{13}{6} \lambda^{3} r^{3}+i \pi  \lambda^{3} r^{3}+\frac{\pi}{2}\gamma^{4}-\frac{\pi}{2}\gamma^{2}+ \nn\\
& \Big(\frac{5}{6}\gamma^{2}+\frac{3}{2}\lambda^{2} r^{2}-\frac{29}{24}\Big) 
\gamma\cot\big(2\pi\gamma\big)+\frac{i}{4}\Big(4\gamma^{2}+4\lambda^{2}r^{2}-3\Big)\lambda r \csc\big(2\pi\gamma\big) \nn\\
&\times \bigg[\Big(e^{2\pi\lambda r}+e^{-2i\pi\gamma}\Big)
\psi\Big(\frac{1}{2}-\gamma+i\lambda r\Big)
-\Big(e^{2\pi\lambda r}+e^{2i\pi\gamma}\Big)\psi\Big(\frac{1}{2}+\gamma+i\lambda r\Big)\bigg],&&
\end{flalign}

\begin{flalign}\label{endi5}
\hspace{0.5cm}
\mathcal{I}_{5} &=-2\Lambda^{2}-\frac{1}{4}\Big(4\gamma^{2}+4\lambda^{2}r^2-1\Big)\log\big(2\Lambda\big)
-\frac{5}{8}-\frac{i \pi}{8}+\frac{\gamma^{2}}{2}+ \frac{\gamma^{2}}{2}i \pi+\frac{3\lambda^{2}r^{2}}{2}+\frac{\lambda^{2}r^{2}}{2}i \pi \nn\\
&+\gamma\lambda r\Big(\cot\big(2\pi\gamma\big)+e^{2\pi\lambda r }\csc\big(2\pi\gamma\big)\Big)+\frac{i}{8}\Big(4\gamma^{2}+4\lambda^{2}r^{2}
-1\Big)\csc\big(2\pi\gamma\big) \nn\\
&\times\bigg[\Big(e^{2\pi\lambda r}+e^{-2i\pi\gamma}\Big)\psi\Big(\frac{1}{2}-\gamma +i\lambda r\Big)-\Big(e^{2\pi\lambda r}+e^ {2i\pi\gamma}\Big)\psi\Big(\frac{1}{2}+\gamma+i\lambda r\Big)\bigg],&&
\end{flalign}

\begin{flalign}\label{endi6}
\hspace{0.5cm}
\mathcal{I}_{6} &=r\lambda \log\big(2\Lambda\big)
-2 \lambda r -\frac{i \pi}{2}\lambda r
-\gamma\Big(\cot\big(2\pi\gamma\big)+e^{2\pi\lambda r }\csc\big(2\pi\gamma\big)\Big)-\frac{i}{2}\lambda r \csc\big(2\pi\gamma\big) \nn\\
&\times\bigg[\Big(e^{2\pi\lambda r}+e^{-2i\pi\gamma}\Big)\psi\Big(\frac{1}{2}-\gamma +i\lambda r\Big)-\Big(e^{2\pi\lambda r}+e^ {2i\pi\gamma}\Big)\psi\Big(\frac{1}{2}+\gamma+i\lambda r\Big)\bigg],&&
\end{flalign}

\end{appendices}

\bibliography{sn-bibliography}% common bib file

\end{document}